\documentclass[sigconf,natbib=true]{acmart}
\AtBeginDocument{%
  \providecommand\BibTeX{{%
    \normalfont B\kern-0.5em{\scshape i\kern-0.25em b}\kern-0.8em\TeX}}}

\usepackage[utf8]{inputenc}
\DeclareUnicodeCharacter{1EAB}{\^a}
\DeclareUnicodeCharacter{1EBF}{\^e}

\copyrightyear{2025}
\acmYear{2025}
\setcopyright{acmlicensed}
\acmConference[SIGIR '25]{Proceedings of the 48th International ACM SIGIR Conference on Research and Development in Information Retrieval}{July 13--18, 2025}{Padua, Italy}
\acmBooktitle{Proceedings of the 48th International ACM SIGIR Conference on Research and Development in Information Retrieval (SIGIR '25), July 13--18, 2025, Padua, Italy}
\acmDOI{10.1145/3726302.3729972}
\acmISBN{979-8-4007-1592-1/2025/07}

\usepackage[utf8]{inputenc} 
\usepackage{amsfonts}       
\usepackage{graphicx}
\usepackage{amsthm}
\usepackage{subfigure}
\usepackage{multirow}
\usepackage{enumitem}
\usepackage{bbding}
\usepackage{colortbl}

\usepackage{algorithmic}
\usepackage{subcaption}
\usepackage{bm}
\usepackage[linesnumbered,ruled,vlined]{algorithm2e}
\usepackage{framed,xcolor}

\newcommand{\indicator}[1]{\mathds{1}}

\newcommand{\wo}{\emph{w/o}\xspace}
\newcommand{\w}{\emph{w/}\xspace}

\definecolor{AIGCcolor}{RGB}{61,169,226}
\definecolor{HGCcolor}{RGB}{247,156,143}
\newcommand{\AIGCcolor}[1]{\textcolor{AIGCcolor}{#1}}
\newcommand{\HGCcolor}[1]{\textcolor{HGCcolor}{#1}}

\newcommand{\hlprimarytab}[1]{\colorbox{white}{#1}}
\newcommand{\better}[1]{{\scriptsize\hlprimarytab{\textcolor{red}{#1}}}}
\newcommand{\worse}[1]{{\scriptsize\hlprimarytab{\textcolor{black}{#1}}}}

\begin{document}

\title{Exploring the Escalation of Source Bias in User, Data, and Recommender System Feedback Loop} 

\author{Yuqi Zhou}
\author{Sunhao Dai}
\affiliation{
  \institution{\mbox{Gaoling School of Artificial Intelligence}\\Renmin University of China}
    \city{Beijing}
  \country{China}
  }
\email{{yuqizhou,sunhaodai}@ruc.edu.cn}

\author{Liang Pang}
\affiliation{
  \institution{CAS Key Laboratory of AI Safety\\Institute of Computing Technology Chinese Academy of Sciences}
    \city{Beijing}
  \country{China}
  }
\email{pangliang@ict.ac.cn}

\author{Gang Wang}
\author{Zhenhua Dong}
\affiliation{
  \institution{Huawei Noah’s Ark Lab	}
    \city{Shenzhen}
  \country{China}
  }
\email{wanggang110@huawei.com}
\email{dongzhenhua@huawei.com}

\author{Jun Xu}
\authornote{Corresponding author}
\affiliation{%
  \institution{\mbox{Gaoling School of Artificial Intelligence}\\Renmin University of China}
 \city{Beijing}
  \country{China}
}
\email{junxu@ruc.edu.cn}

\author{Ji-Rong Wen}
\affiliation{
  \institution{\mbox{Gaoling School of Artificial Intelligence}\\Renmin University of China}
    \city{Beijing}
  \country{China}
  }
\email{jrwen@ruc.edu.cn}

\renewcommand{\authors}{Yuqi Zhou, Sunhao Dai, Liang Pang, Gang Wang, Zhenhua Dong, Jun Xu, Ji-Rong Wen}
\renewcommand{\shortauthors}{Yuqi Zhou et al.}


\begin{abstract}

\end{abstract}



\begin{abstract}
Recommender systems are essential for information access, allowing users to present their content for recommendation. With the rise of large language models (LLMs), AI-generated content (AIGC), primarily in the form of text, has become a central part of the content ecosystem. As AIGC becomes increasingly prevalent, it is important to understand how it affects the performance and dynamics of recommender systems. To this end, we construct an environment that incorporates AIGC to explore its short-term impact. The results from popular sequential recommendation models reveal that \textbf{AIGC are ranked higher in the recommender system}, reflecting the phenomenon of source bias~\cite{dai2024neural,xu2023ai}. To further explore the long-term impact of AIGC, we introduce a feedback loop with realistic simulators. The results show that the model’s preference for AIGC increases as the user clicks on AIGC rises and the model trains on simulated click data. This leads to two issues: In the short term, bias toward AIGC encourages LLM-based content creation, increasing AIGC content, and causing unfair traffic distribution. From a long-term perspective, our experiments also show that when AIGC dominates the content ecosystem after a feedback loop, it can lead to a decline in recommendation performance. To address these issues, we propose a debiasing method based on L1-loss optimization to maintain long-term content ecosystem balance. In a real-world environment with AIGC generated by mainstream LLMs, our method ensures a balance between AIGC and human-generated content in the ecosystem. The code and dataset are available at \textcolor{blue}{\url{https://github.com/Yuqi-Zhou/Rec_SourceBias}}.

\end{abstract}

\ccsdesc[500]{Information systems~Recommender systems}
\keywords{Source Bias, AI-Generated Content, Large Language Model}
\maketitle

\section{Introduction}

\begin{figure}[t]  
    \centering    
    \includegraphics[width=0.9\linewidth]{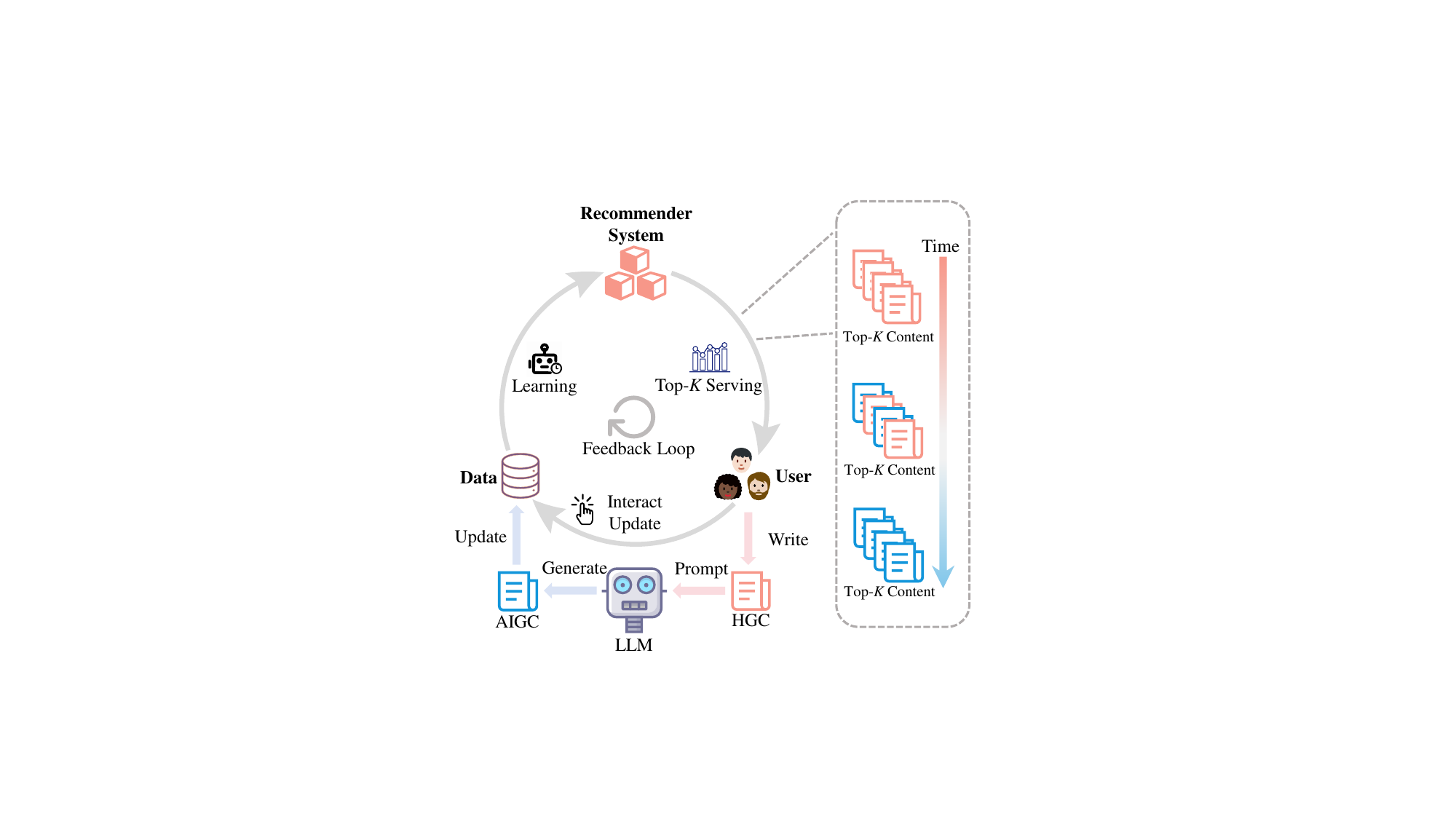}
    \caption{Model preference grows over time with the feedback loop of humans, data, and the recommender system. The \HGCcolor{red} color is used for HGC icon and the \AIGCcolor{blue} color is used for AIGC icon. The subsequent figures use the same color scheme.}
    \label{fig:recommendation_framework}  
\end{figure}

\begin{figure*}[t]  
    \centering    
    \includegraphics[width=0.9\linewidth]{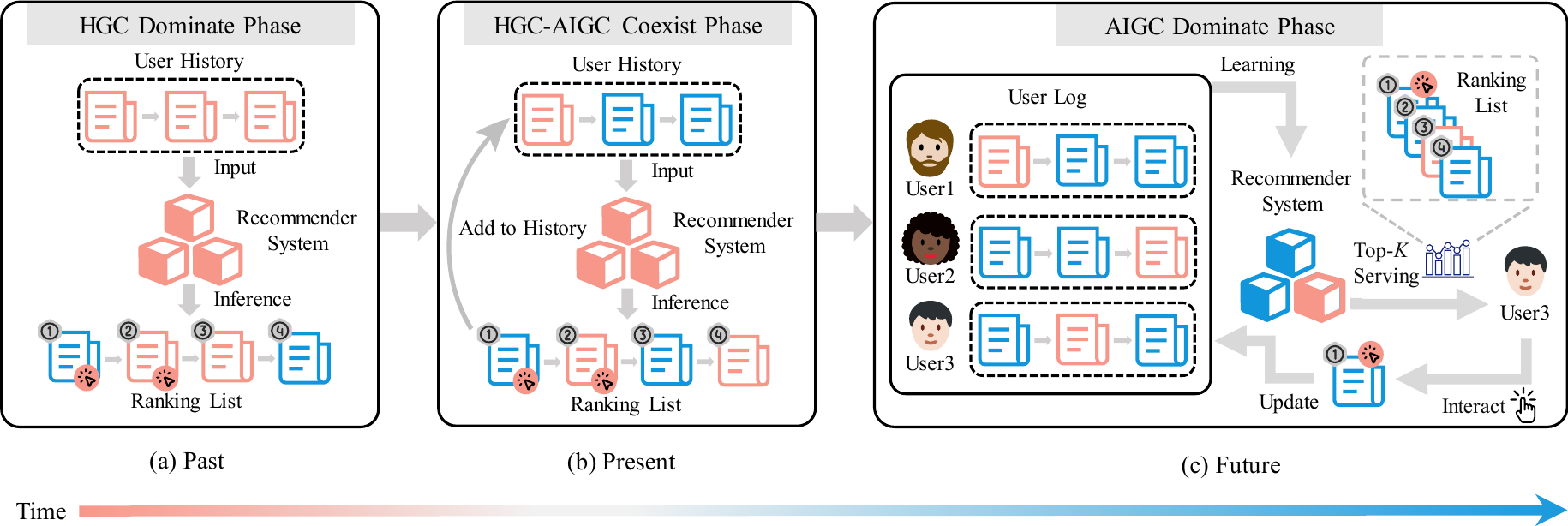}
    \caption{Three phases occur during the integration of AIGC into the recommendation content ecosystem: HGC dominate phase, HGC-AIGC coexist phase, and AIGC dominate phase. (1) The HGC dominate phase is a past period when AIGC has just flooded into the recommender systems and only influence the candidate list. (2) The HGC-AIGC coexist phase is a present period where the recommendation model's inputs $s$ contain an increasing number of AIGC. (3) The AIGC dominate phase is a future period during which AIGC influences each stage of the feedback loop.}
    \label{fig:feedback_loop}  
\end{figure*}

The development of Large Language Models (LLMs) is advancing rapidly~\cite{zhao2023survey}, demonstrating strong capabilities and performing excellently in many text generation tasks, such as machine translation~\cite{lai2023chatgpt}, summarization~\cite{zhang2023extractive}, and complex question answering~\cite{tan2023evaluation,xu2024search}. Due to the cost-effectiveness, high quality, and speed of generating AIGC compared to Human Generated Content (HGC), an increasing volume of online content is being produced by various LLMs and ~\citet{schick_deep_2020} suggests that the synthetic content could dominate up to 90\% of the Internet. This shift is altering the existing content creation paradigm and resulting in a prevalence of AIGC on the internet~\cite{dai2024bias, dai2025unifying}.

When AIGC floods into the internet, these contents will be disseminated by the current information retrieval systems, especially recommender systems, which play a central role in shaping users' online experiences. However, the impact of this rapidly growing AIGC content on current and future recommender systems has yet to be explored. Therefore, an important research question emerges: \textbf{RQ1: What short-term impacts will the influx of AIGC have on recommender systems?} This research primarily focuses on AIGC in the form of high-quality \textbf{text} generated by LLMs, which are increasingly prevalent on the internet. Unlike other modalities such as images, LLM-generated text is harder to distinguish, potentially introducing more subtle biases. In the recommender system, feedback data from user interactions with LLM-generated text is reused to update recommendation models, forming a feedback loop involving users, data, and the system as shown in Figure \ref{fig:recommendation_framework}.  With the continuous increase in AIGC, they will gradually influence various stages of the feedback loop, raising another research question: \textbf{RQ2: What long-term impacts will arise when AIGC further participates in the feedback loop?}

To explore the two research questions, we first examine AIGC’s impact on recommender systems across three phases: HGC dominate phase, HGC-AIGC coexist phase, and AIGC dominate phase, as shown in Figure~\ref{fig:feedback_loop}. These phases respectively correspond to the past, present, and future, representing the influence of different stages of AIGC flooding on recommender systems. In the HGC-dominated phase, AIGC starts influencing the candidate set and Top-$K$ results within the feedback loop. During the HGC-AIGC coexist phase, AIGC further affects users’ histories via interaction, potentially amplifying earlier effects. In the AIGC dominate phase, AIGC prevails in training, likely intensifying these influences.

To answer RQ1, we construct evaluation datasets from three domains in Amazon's product dataset by prompting LLMs to rewrite product descriptions into AIGC copies~\cite{dai2024neural}. We then evaluate popular recommendation models using mixed candidate sets of HGC and AIGC items during the \textbf{HGC dominate phase}. Results show that models often rank AIGC copies higher, even when their semantics match the original HGC. This suggests product traffic can be boosted by rewriting descriptions with LLMs, while also highlighting risks—such as enabling malicious users to spread fake news~\cite{zhou2020survey} via LLM-generated content.

For RQ2, we conduct experiments by injecting AIGC into users’ interaction histories during the \textbf{HGC-AIGC coexist phase}, and into model training data during the \textbf{AIGC-dominated phase}, within a feedback loop scenario. Results from four widely used click models show that both user behaviors and model updates on polluted data increasingly favor AIGC items. Consequently, AIGC content is ranked progressively higher, eventually reaching top positions. Moreover, as the loop progresses and AIGC proportion rises, we observe a corresponding decline in the recommendation model's performance. This further underscores the detrimental impact of excessive AIGC on recommender systems.

Based on the above findings, content creators may rewrite all descriptions to gain higher rankings, creating unfairness for other providers. Moreover, due to hallucinations~\cite{huang2023survey}, LLM-generated texts may contain inaccuracies, harming user experience. Prior studies and our results further suggest that training on AIGC clicked by users can cause model collapse~\cite{shumailov2023model,alemohammad2023self,briesch2023large}, degrading recommendation performance. In sum, AIGC’s dominance in recommender systems poses long-term risks to content fairness, user experience, and model quality. This calls for disrupting AIGC propagation in the feedback loop, leading to a new research question:
\textbf{RQ3: How can the model maintain consistent preferences for both HGC and AIGC in the feedback loop?}

To answer RQ3, we first examine prior debiasing methods~\cite{dai2024neural,xu2023ai,cocktail} and find they fail in feedback loop scenarios, unable to maintain long-term system balance. To address this, we propose a black-box debiasing method that preserves model neutrality toward both HGC and AIGC. Our approach prompts LLMs to uniformly rewrite all training data to get AIGC copies, avoiding the need to distinguish between sources. We then apply an L1 loss to constrain outputs of the item and history encoders, ensuring semantically similar HGC and AIGC are mapped to aligned embeddings. Experiments show our method effectively reduces bias and maintains prediction neutrality across varying AIGC proportions.

The major contributions of this paper are summarized as follows:

($1$) We find that LLM-generated text descriptions can be ranked higher in recommender systems. 

($2$) We uncover that the recommendation model's preference for AIGC is gradually amplified in the feedback loop, with AIGC sequentially affecting data, users, and recommender systems. 

($3$) We propose a debiasing method that can effectively alleviate preference during the feedback loop by aligning the item and user embedding spaces, thereby balancing the content ecosystem.

\section{Preliminaries}\label{sec:framework}
In this section, we formulate the recommendation problem, introduce three stages of AIGC flooding into recommender systems, and explore the role of the feedback loop in propagating source bias.

\subsection{Recommendation Problem Formulation}
Assume that we have a set of items $\mathcal{I}$ and a set of user interaction sequences $\mathcal{S}$, where $i \in \mathcal{I}$ denotes an item and $s \in \mathcal{S}$ denotes an interaction sequence. The numbers of items and sequences are denoted as $|\mathcal{I}|$ and $|\mathcal{S}|$, respectively. Generally, the interaction sequence $s$ is chronologically ordered with items: $\{i_1,\cdots,i_n\}$, where $n$ is the number of interactions and $i_t$ is the $t$-th item with which the user has interacted. For convenience, we use $s_t$ to denote the subsequence, \textit{i.e.}, $s_t=\{i_1,\cdots,i_t\}$, where $1 \leq t < n$.

Based on the above notations, we now define the task of recommendation. Formally, given the history interaction sequence of a user $s_t=\{i_1,\cdots, i_t\}$, the goal of recommendation is to train a recommendation model $f_\theta$ parameterized by $\theta$. The model $f_\theta$ is used to predict the next item $i_{t+1}$ the user is likely to interact with at the $(t+1)$-th step.

\subsection{Three Phases Involving AIGC Content}

After AIGC integrates into the content ecosystem of recommender systems, it will gradually impact the three processes of the feedback loop over time: Top-$K$ serving, users' interaction, and model training. We divide the impact of AIGC on recommender systems by feedback loop over time into three phases: HGC Dominate, HGC-AIGC Coexist, and AIGC Dominate. Each phase corresponds to a real-world scenario representing the past, present, and future.

\textbf{HGC Dominate Phase:}
With the widespread use of LLMs and the popularization of AIGC on the internet, it is easy for HGC to have corresponding AIGC copies or even be directly generated by LLMs. Thus, the items selected for the recommendation model's Top-$K$ ranking are a combination of HGC and AIGC. In the HGC dominate phase, the research question aims to validate whether the recommendation models will rank AIGC at a higher position, a phenomenon known as source bias~\cite{dai2024neural, xu2023ai, cocktail}.

\textbf{HGC-AIGC Coexist Phase:} With the increasing proliferation of LLMs and AIGC on the Internet, the presence of AIGC in users' recommendation candidate lists will rapidly grow. These contents will be interacted with users and added to their interaction sequences, which will be used as input for recommendation models later. In the HGC-AIGC coexist phase, the research question is whether the model's preference for AIGC will be amplified when AIGC interacted with users is added to users' interaction sequence. 

\textbf{AIGC Dominate Phase:} In the future, with the decreasing cost and increasing accessibility of LLMs, AIGC will dominate the ecosystem of recommender systems. Furthermore, AIGC will influence any stage of the feedback loop, namely Top-$K$ serving, users' interaction, and model training in Figure~\ref{fig:recommendation_framework}. 
In other words, AIGC will pollute candidate list $\mathcal{I}$, users' interaction history sequence $s$, and the model's training data $\mathcal{S}$. Furthermore, within the iterative feedback loop, recommendation models undergo training on data $\mathcal{S}$ containing AIGC. In the AIGC dominate phase, the research question is whether the preference will be amplified when recommendation models are further trained on polluted data.

In conclusion, the integration of AIGC into the recommender system will impact various aspects, such as the candidate item set, users' interactions, and data used for model training. Based on the affected aspects, the evolution of the recommender system will progressively exhibit three phases: HGC Dominate, HGC-AIGC Coexist, and AIGC Dominate. We will explore the changes in preference across these three phases to answer RQ1 and RQ2.

\section{Source Bias in Recommender Systems}
In this section, we first introduce the experimental settings in Section~\ref{sec:settings} and then provide the data construction process and verify the AIGC quality through human evaluation in Section~\ref{sec:data_construction}. In Section~\ref{sec:sec_loop1}, we validate the existence of source bias in recommender systems during the HGC dominate phase. In Section~\ref{sec:sec_loop2} and Section~\ref{sec:sec_loop3}, we verify that source bias is amplified in the feedback loop due to users' interaction behavior and the model training process.

\subsection{Experimental Settings}\label{sec:settings}
\subsubsection{Datasets}
Our training and evaluation are conducted on a series of real-world datasets (Amazon~\cite{mcauley2015image}), comprising large corpora of product reviews and descriptions obtained from Amazon.com. Top-level product categories are treated as separate datasets, and we focus on three categories: ``Health'', ``Beauty'', and ``Sports''. We use the textual descriptions of products that users have commented on as input to predict which product the user might review next. Due to the low quality of short text rewriting\footnote{LLM frequently expands short texts during the rewriting process, leading to semantic }, we exclude items with descriptions containing fewer than 20 words from the training set to maintain training stability. We sort the data based on the review time of the target item and split it into training and testing sets in a 7:3 ratio. In the training dataset, we exclude users and items with fewer than five interactions and randomly select 4 negative items from the entire set for each product reviewed by users. The statistics of datasets after processing are shown in Table\ref{tab: data_statistic}.
\begin{table}[t]
\caption{Statistics of the experimental datasets.}
\vspace{-10pt}
\label{tab: data_statistic}
\begin{tabular}{lccc}
\toprule
Dataset  & Health & Beauty & Sports   \\
\midrule
\# Users           & 18,036   & 11,391  &    16,639  \\
\# Items           &  13,972   & 11,897  &    13,089 \\
\# Click Behaviors &  346,355   & 198,502  &  296,337  \\         
\bottomrule
\end{tabular}
\vspace{-6pt}
\end{table}

\subsubsection{Recommendation Models}
For our main experiments, we select four representative models: BERT4Rec~\cite{sun2019bert4rec}, SASRec~\cite{kang2018self}, GRU4Rec~\cite{hidasi2015session}, LRURec~\cite{yue2024linear}. These models are enhanced by various pre-trained language models, including BERT~\cite{kenton2019bert} and RoBERTa~\cite{liu2019roberta}. Our focus on sequential recommendation models is motivated by their widespread use in current industrial recommender systems. We input the product's textual description into these pretrained models and use the average pooled embedding of the outputs from the pretrained models as the item embedding. Item embedding is then used as the input for the four sequence recommendation models mentioned above. We use the \texttt{bert-base-uncased} checkpoint for BERT and the \texttt{roberta-base} checkpoint for RoBERTa.

\subsubsection{Evaluation Metrics}

To evaluate the ranking performance of the recommendation models, we compute the Top-$K$ Normalized Discounted Cumulative Gain (NDCG@$K$) and Mean Average Precision (MAP@$K$) separately for HGC and AIGC items, where $K \in \{3,5\}$.
To further measure the recommendation models' preferences for different source texts, the candidates during testing are divided into two parts: one part consists of original HGC, and the other part consists of copies of AIGC. 
To get a simple and efficient measuring way, we utilize the relative percentage difference~\cite{dai2024neural, xu2023ai, cocktail}: 
\begin{align}
 \text{Relative $\Delta$} = \frac{\texttt{Metric}_\text{HGC} - \texttt{Metric}_\text{AIGC}}{(\texttt{Metric}_\text{HGC} +\texttt{Metric}_\text{AIGC}) / 2} \times 100\%,
\end{align}
where $\texttt{Metric}_{\text{HGC}}$ and $\texttt{Metric}_{\text{AIGC}}$ are calculated on the same candidate set comprising both HGC and AIGC. For a given metric (either NDCG@$K$ or MAP@$K$), when measuring the metric for one data source, we set the labels of the other data source to 0. Relative $\Delta > 0$ indicates a preference of the recommendation models towards HGC, while Relative $\Delta < 0$ indicates a preference towards AIGC. The greater the absolute value of Relative $\Delta$, the stronger the preference recommendation model for AIGC or HGC.

\subsubsection{Experimental Details}
To ensure computational efficiency, all pre-trained language models are frozen. All recommendation models are trained for $5$ epochs, and the best-performing model on the development set is selected for testing on the test set.
The batch size is set to $128$, the learning rate is set to $1e\text{-}3$, and the Adam optimizer is used for training. The dimension of item vectors is set to $768$, and all score calculations utilize the dot function. The text input to the model is truncated to 512 tokens, and the user's historical sequence is limited to 10 interactions. To ensure reproducibility, we run each experiment with five different seeds and report the averaged results.

\subsection{AIGC Data Construction and Verification}\label{sec:data_construction}
\subsubsection{Data Construction}
Following the setting in previous works~\cite{dai2024neural, xu2023ai, cocktail}, we reconstruct the dataset from Amazon to evaluate source bias in recommender systems. For each item $i \in \mathcal{I}$, we utilize the same rewriting prompt ``\textit{Please rewrite the following text: \{\{human-written text\}\}}'' to empower LLMs to produce text without extra constraints, all the while upholding semantic equivalence to the initial HGC. Specifically, we chose some popular LLMs ChatGPT (\textit{i.e., }\texttt{gpt-3.5-turbo-0613}), Llama (\textit{i.e., }\texttt{llama-2-7b-chat})~\cite{touvron2023llama}, Mistral (\textit{i.e., }\texttt{Mistral-7B-Instruct-v0.2})~\cite{jiang2023mistral}, and Gemini-pro (\textit{i.e., }\texttt{Gemini 1.5 Pro})~\cite{team2023gemini} to rewrite each seed HGC, as these LLMs are the most widely used. The temperature of all LLMs for generation is set at $0.2$ and the maximum generation length is $256$.\footnote{The examples of rewritten text can be found in the supporting materials.}

After rewriting, we can obtain HGC data and the corresponding AIGC data for each dataset.
Formally, we have two sets of items denoted by $\mathcal{I}^{H}$ and $\mathcal{I}^{G}$, respectively. Here, $i^H \in \mathcal{I}^H$ represents an item written by a human, while $i^G \in \mathcal{I}^G$ represents an item generated by LLMs. Each item $i^H$ has its corresponding AIGC copy $i^G \in \mathcal{I}^G$. In the LLMs era, the task of recommendation is to predict the next item $i_{t+1}$ the user is likely to interact with from a mixed set of items $\mathcal{I}=\mathcal{I}^H \cup \mathcal{I}^G$, rather than just $\mathcal{I}^H$.

\begin{figure}[t]  
    \centering    
    \includegraphics[width=0.8\linewidth]{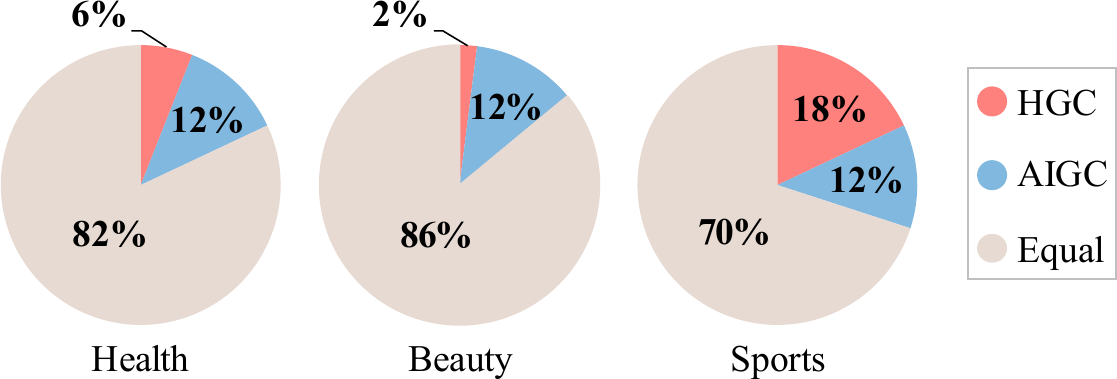}
    \vspace{-5pt}
    \caption{Quality verification of the constructed datasets through human evaluation.}
    \label{fig:human_evaluation}  
\end{figure}

\begin{table*}
\caption{Performance comparison of recommendation models based on BERT and RoBERTa for mixed HGC and AIGC item sets on the Health, Beauty, and Sports dataset. Relative $\Delta < 0$ indicates that the recommendation models rank AIGC higher than HGC, while Relative $\Delta > 0$ indicates that the models rank HGC higher than AIGC. The absolute value of Relative $\Delta$ indicates the degree of bias, with a larger value representing a stronger bias. Unless otherwise stated, AIGC will be generated using ChatGPT with BERT as the encoder model, and Relative $\Delta$ is calculated based on NDCG@$5$.}
\vspace{-5pt}
\centering
\label{tab:bias_comparison}
\resizebox{1.0\textwidth}{!}{
\begin{tabular}{cllcccccccccccc}
\hline\hline
\multirow{2}{*}{PLM}      & \multirow{2}{*}{Model}      & \multirow{2}{*}{Corpus}       & \multicolumn{4}{c}{Health} & \multicolumn{4}{c}{Beauty} & \multicolumn{4}{c}{Sports} \\
\cmidrule(lr){4-7} \cmidrule(lr){8-11} \cmidrule(lr){12-15}
&          &            & NDCG@3   & NDCG@5 & MAP@3 & MAP@5 & NDCG@3   & NDCG@5 & MAP@3 & MAP@5 & NDCG@3   & NDCG@5 & MAP@3 & MAP@5 \\
\hline
 \multirow{12}{*}{BERT} & \multirow{3}{*}{GRU4Rec} & Human-Written & 32.77 & 41.18 & 28.39 & 33.06 & 27.92 & 36.88 & 23.80 & 28.79 & 31.80 & 41.12 & 26.75 & 31.97 \\
    &  & LLM-Generated & 41.28 & 48.74 & 36.92 & 41.03 & 44.21 & 52.41 & 39.54 & 44.09 & 51.73 & 58.92 & 47.03 & 51.00 \\
    & & Relative $\Delta$ & \textcolor{red}{-22.99} & \textcolor{red}{-16.80} & \textcolor{red}{-26.13} & \textcolor{red}{-21.51} & \textcolor{red}{-45.18} & \textcolor{red}{-34.78} & \textcolor{red}{-49.71} & \textcolor{red}{-42.01} &  \textcolor{red}{-47.74} & \textcolor{red}{-35.60} & \textcolor{red}{-54.98} & \textcolor{red}{-45.87} \\
\cline{2-15}
 & \multirow{3}{*}{SASRec} & Human-Written & 24.47 & 32.74 & 20.54 & 25.11 & 23.80 & 32.27 & 20.32 & 25.00  & 25.45 & 34.91 & 21.39 & 26.62 \\ 
    &  & LLM-Generated &  39.82 & 47.60 & 35.80 & 40.10 & 34.51 & 43.38 & 30.28 & 35.18 & 44.56 & 52.73 & 39.95 & 44.48 \\
    & & Relative $\Delta$ & \textcolor{red}{-47.76} & \textcolor{red}{-36.99} & \textcolor{red}{-54.16} & \textcolor{red}{-45.96} & \textcolor{red}{-36.73} & \textcolor{red}{-29.38} & \textcolor{red}{-39.34} & \textcolor{red}{-33.85} & \textcolor{red}{-54.58} & \textcolor{red}{-40.68} & \textcolor{red}{-60.50} & \textcolor{red}{-50.24} \\
\cline{2-15}
& \multirow{3}{*}{BERT4Rec} & Human-Written &   26.49 & 35.22 & 22.88 & 27.70 & 21.53 & 30.31 & 18.42 & 23.27 & 26.00 & 35.54 & 22.06 & 27.37 \\ 
    &  & LLM-Generated &   32.89 & 40.97 & 28.67 & 33.12 & 35.51 & 43.52 & 31.25 & 35.67 & 40.46 & 49.12 & 36.17 & 40.95 \\ 
    & & Relative $\Delta$ &  \textcolor{red}{-21.57} & \textcolor{red}{-15.09} & \textcolor{red}{-22.47} & \textcolor{red}{-17.85} & \textcolor{red}{-49.00} & \textcolor{red}{-35.81} & \textcolor{red}{-51.66} & \textcolor{red}{-42.09} & \textcolor{red}{-43.49} & \textcolor{red}{-32.06} & \textcolor{red}{-48.46} & \textcolor{red}{-39.77} \\
\cline{2-15}
 & \multirow{3}{*}{LRURec} & Human-Written & 34.22 & 42.13 & 30.13 & 34.51 & 31.30 & 39.26 & 27.26 & 31.66 & 34.12 & 42.31 & 29.55 & 34.09 \\
    &  & LLM-Generated & 32.29 & 40.22 & 28.33 & 32.69 & 38.30 & 46.24 & 33.84 & 38.24 & 40.84 & 49.40 & 36.10 & 40.86 \\ 
    & & Relative $\Delta$ & \textcolor{red}{-20.13} & \textcolor{black}{5.80} & \textcolor{black}{4.65} & \textcolor{black}{6.16} & \textcolor{black}{5.43} & \textcolor{red}{-16.35} & \textcolor{red}{-21.54} & \textcolor{red}{-18.82} &  \textcolor{red}{-17.92} & \textcolor{red}{-15.46} & \textcolor{red}{-19.93} & \textcolor{red}{-18.05} \\
\hline
\multirow{12}{*}{RoBERTa} 
 & \multirow{3}{*}{GRU4Rec} & Human-Written & 30.96 & 39.01 & 26.52 & 31.01 & 36.64 & 45.07 & 32.35 & 36.98 & 40.54 & 48.72 & 35.88 & 40.43 \\ 
    &  & LLM-Generated & 44.10 & 50.98 & 39.48 & 43.26 & 34.61 & 43.58 & 30.00 & 34.98 & 43.43 & 50.82 & 38.61 & 42.72 \\ 
    & & Relative $\Delta$ & \textcolor{red}{-35.01} & \textcolor{red}{-26.60} & \textcolor{red}{-39.27} & \textcolor{red}{-32.97} &  \textcolor{black}{5.69} & \textcolor{black}{3.36} & \textcolor{black}{7.55} & \textcolor{black}{5.55}  & \textcolor{red}{-6.88} & \textcolor{red}{-4.20} & \textcolor{red}{-7.33} & \textcolor{red}{-5.51} \\
\cline{2-15}
 & \multirow{3}{*}{SASRec} & Human-Written & 24.54 & 32.53 & 20.81 & 25.25 & 20.74 & 29.56 & 17.74 & 22.59 & 17.79 & 24.70 & 15.31 & 19.12 \\ 
    &  & LLM-Generated & 38.63 & 46.76 & 34.32 & 38.78 & 27.87 & 36.93 & 24.24 & 29.26 & 26.57 & 35.65 & 23.00 & 28.00 \\ 
    & & Relative $\Delta$ & \textcolor{red}{-44.62} & \textcolor{red}{-35.91} & \textcolor{red}{-49.02} & \textcolor{red}{-42.26} & \textcolor{red}{-29.32} & \textcolor{red}{-22.17} & \textcolor{red}{-30.94} & \textcolor{red}{-25.70} & \textcolor{red}{-39.57} & \textcolor{red}{-36.30} & \textcolor{red}{-40.15} & \textcolor{red}{-37.69} \\
\cline{2-15}
& \multirow{3}{*}{BERT4Rec} & Human-Written & 29.37 & 37.60 & 25.40 & 29.97 & 26.29 & 34.86 & 22.84 & 27.59 & 31.14 & 39.98 & 27.15 & 32.05 \\ 
    &  & LLM-Generated & 40.02 & 47.20 & 35.55 & 39.52 & 29.95 & 38.28 & 25.81 & 30.40 & 39.26 & 47.96 & 34.60 & 39.45 \\
    & & Relative $\Delta$ & \textcolor{red}{-30.68} & \textcolor{red}{-22.65} & \textcolor{red}{-33.30} & \textcolor{red}{-27.48} & \textcolor{red}{-13.02} & \textcolor{red}{-9.35} & \textcolor{red}{-12.22} & \textcolor{red}{-9.68} & \textcolor{red}{-23.06} & \textcolor{red}{-18.14} & \textcolor{red}{-24.16} & \textcolor{red}{-20.68} \\
\cline{2-15}
 & \multirow{3}{*}{LRURec} & Human-Written & 25.35 & 34.33 & 21.19 & 26.15 & 33.06 & 40.23 & 29.40 & 33.36 & 30.24 & 39.49 & 26.30 & 31.42 \\  
    &  & LLM-Generated & 44.21 & 51.24 & 39.71 & 43.59 &  30.37 & 39.95 & 25.95 & 31.26 & 39.32 & 48.00 & 34.55 & 39.36 \\ 
    & & Relative $\Delta$ & \textcolor{red}{-54.21} & \textcolor{red}{-39.51} & \textcolor{red}{-60.80} & \textcolor{red}{-50.01} &  \textcolor{black}{8.46} & \textcolor{black}{0.70} & \textcolor{black}{12.46} & \textcolor{black}{6.50} & \textcolor{red}{-26.11} & \textcolor{red}{-19.47} & \textcolor{red}{-27.12} & \textcolor{red}{-22.44} \\
\cline{2-15}

\hline\hline
\end{tabular}
}
\end{table*}
\subsubsection{Human Evaluation}
To validate that the rewritten data does not affect users' interaction behaviors, we conduct a human evaluation study by sampling $50$ triples from the Health, Beauty, and Sports, respectively. For each domain, we recruit three colleagues for data annotation. Each human annotator is asked to indicate which item they would be more inclined to purchase based on the textual description of products in the browsing purchase history, with options being ``Human items'', ``LLM items'' and ``Equal''. Each triple is annotated by at least three annotators, and the votes determine the final label. The evaluation results in Figure~\ref{fig:human_evaluation} demonstrate the consistency of humans' interaction behaviors on HGC and AIGC, providing reliable assurance for the evaluation and analysis of source bias.

\subsection{Preference in HGC Dominate Phase}\label{sec:sec_loop1}

In this subsection, we examine the recommendation models during the HGC dominate phase, aiming to explore whether AIGC will be ranked higher. We train recommendation models on each dataset with items from $\mathcal{I}^H$ and test the model's performance on candidate items from $\mathcal{I}^H \cup \mathcal{I}^G$. As shown in Table~\ref{tab:bias_comparison}, it can be observed that most recommendation models exhibit preference for AIGC in terms of metrics such as NDCG@$K$ and MAP@$K$. An important point is that the higher metric on LLM-Generated compared to Human-Written does not imply better ranking performance on AIGC. When measuring, the candidate set is $I^H \cup I^G$, with each HGC having a AIGC-copy. For Human-Written, AIGC’s positive samples are treated as negative, focusing only on HGC's positive item. Thus, the higher score on LLM-Generated only reflects HGC have a lower ranking score than its AIGC-copy.

To verify the widespread presence of preference in recommender systems, we test recommendation models on AIGC generated by more popular LLMs such as Llama, Mistral, and Gemini-Pro. The results in Table~\ref{tab:llm_bias_comparison} indicate the varying degrees of preference on AIGC generated by different LLMs, confirming the prevalence and significance of preference. Furthermore, ChatGPT demonstrates a smaller preference compared to other LLMs, likely due to its better alignment with human behavior during pre-training.

\begin{center}
\fcolorbox{gray!60}{gray!20}{
  \parbox{0.95\linewidth}{
    \textbf{Finding 1:} During the HGC dominate phase, various recommendation models based on different PLMs tend to show a preference for AIGC generated by various LLMs across three datasets from diverse domains.
  }
}
\end{center}
\begin{table*}
\caption{Relative $\Delta$ of recommendation models with AIGC copies generated by ChatGPT, Llama2, Mistral, and Gemini-Pro.}
\vspace{-5pt}
\centering
\label{tab:llm_bias_comparison}
\resizebox{1.0\textwidth}{!}{
\begin{tabular}{clcccccccccccc}
\hline\hline
\multirow{2}{*}{PLM} &  \multirow{2}{*}{Model}         & \multicolumn{4}{c}{Health} & \multicolumn{4}{c}{Beauty} & \multicolumn{4}{c}{Sports} \\
\cmidrule(lr){3-6} \cmidrule(lr){7-10} \cmidrule(lr){11-14}
&              & ChatGPT  & Llama2 & Mistral & Gemini-Pro & ChatGPT  & Llama2 & Mistral & Gemini-Pro & ChatGPT  & Llama2 & Mistral & Gemini-Pro \\
\hline
\multirow{4}{*}{BERT} & GRU4Rec & \textcolor{red}{-16.80} & - & \textcolor{red}{-18.3} & \textcolor{red}{-23.7} & \textcolor{red}{-34.78} & \textcolor{red}{-37.91} & \textcolor{red}{-33.83} & \textcolor{red}{-47.62} &  \textcolor{red}{-35.60} & \textcolor{red}{-24.14} & \textcolor{red}{-49.58} & \textcolor{red}{-43.23} \\
          & SASRec & \textcolor{red}{-36.99} & - & \textcolor{red}{-46.07} & \textcolor{red}{-50.66} & \textcolor{red}{-29.38} & \textcolor{red}{-36.57} & \textcolor{red}{-30.82} & \textcolor{red}{-62.06} &  \textcolor{red}{-40.68} & \textcolor{red}{-31.7} & \textcolor{red}{-53.36} & \textcolor{red}{-56.17} \\
        & BERT4Rec  &  \textcolor{red}{-15.09} & - & \textcolor{red}{-9.093} & \textcolor{red}{-22.69} & \textcolor{red}{-35.81} & \textcolor{red}{-23.94} & \textcolor{red}{-18.75} & \textcolor{red}{-35.93} & \textcolor{red}{-32.06} & \textcolor{red}{-2.134} & \textcolor{red}{-38.23} & \textcolor{red}{-46.35}  \\
          & LRURec & \textcolor{black}{4.65} & - & \textcolor{red}{-10.83} & \textcolor{red}{-11.70} & \textcolor{red}{-16.35} & \textcolor{red}{-27.11} & \textcolor{red}{-13.32} & \textcolor{red}{-31.57} &  \textcolor{red}{-15.46} & \textcolor{red}{-44.10} & \textcolor{red}{-35.7} & \textcolor{red}{-36.12}
 \\  
\hline        
\multirow{4}{*}{RoBERTa} & GRU4Rec & \textcolor{red}{-26.60} & - & \textcolor{red}{-20.21} & \textcolor{red}{-27.27} & \textcolor{black}{3.36} & \textcolor{red}{-28.39} & \textcolor{red}{-16.50} & \textcolor{red}{-35.77} &  \textcolor{red}{-4.20} & \textcolor{red}{-2.685} & \textcolor{red}{-19.71} & \textcolor{red}{-19.32} \\
          & SASRec & \textcolor{red}{-35.91} & - & \textcolor{red}{-41.5} & \textcolor{red}{-52.13} &  \textcolor{red}{-22.17} & \textcolor{red}{-42.13} & \textcolor{red}{-12.58} & \textcolor{red}{-25.95} & \textcolor{red}{-36.30} & \textcolor{red}{-81.06} & \textcolor{red}{-58.85} & \textcolor{red}{-54.86} \\  
          & BERT4Rec & \textcolor{red}{-22.65} & - & \textcolor{red}{-27.44} & \textcolor{red}{-40.10} & \textcolor{red}{-9.35} & \textcolor{red}{-44.29} & \textcolor{red}{-20.79} & \textcolor{red}{-52.88} & \textcolor{red}{-18.14} & \textcolor{red}{-18.79} & \textcolor{red}{-30.30} & \textcolor{red}{-41.18} \\  
          & LRURec & \textcolor{red}{-39.51} & - & \textcolor{red}{-41.66} & \textcolor{red}{-49.38} & \textcolor{black}{0.70} & \textcolor{red}{-13.73} & \textcolor{red}{-7.91} & \textcolor{red}{-23.23} &  \textcolor{red}{-19.47} & \textcolor{red}{-32.55} & \textcolor{red}{-39.02} & \textcolor{red}{-37.94} \\ 
\hline\hline
\multicolumn{14}{l}{Note: We omit the result for Health dataset
 as Llama2 refuses to rewrite $97.7\%$ of the product description due to that Health contains sensitive information.}
\end{tabular}
}
\end{table*}

\begin{figure*}[t]
  \subfigure[Health]
    {
    \includegraphics[width=0.65\columnwidth]{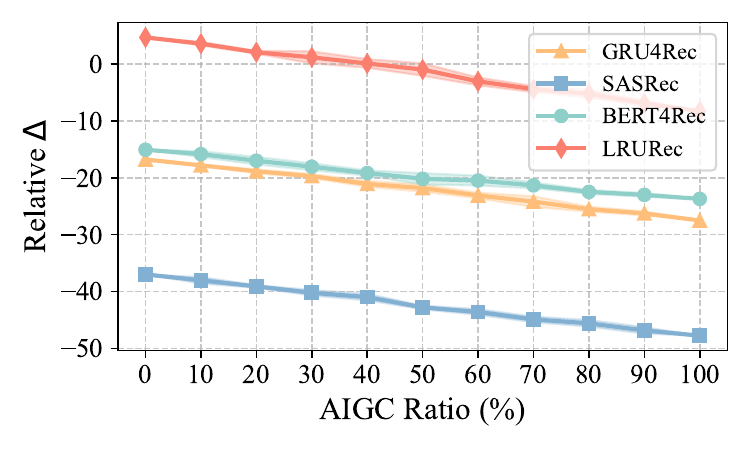}
    \label{fig:loop2_Amazon_Health}
   }
  \subfigure[Beauty]
    {
    \includegraphics[width=0.65\columnwidth]{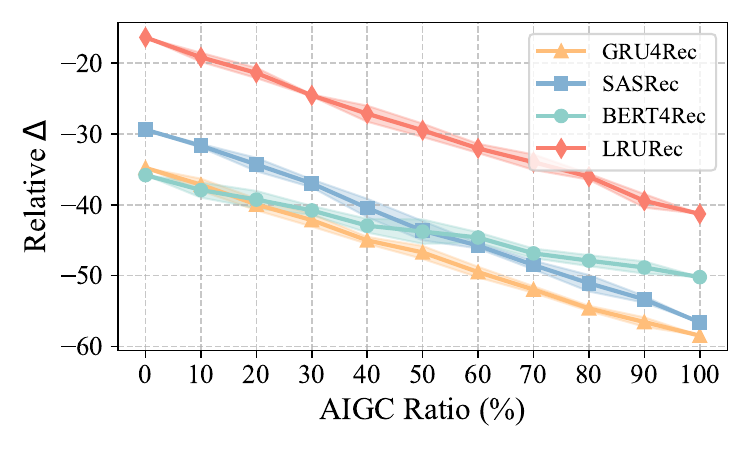}
    \label{fig:loop2_Amazon_Beauty}
   }
  \subfigure[Sports]
    {
    \includegraphics[width=0.65\columnwidth]{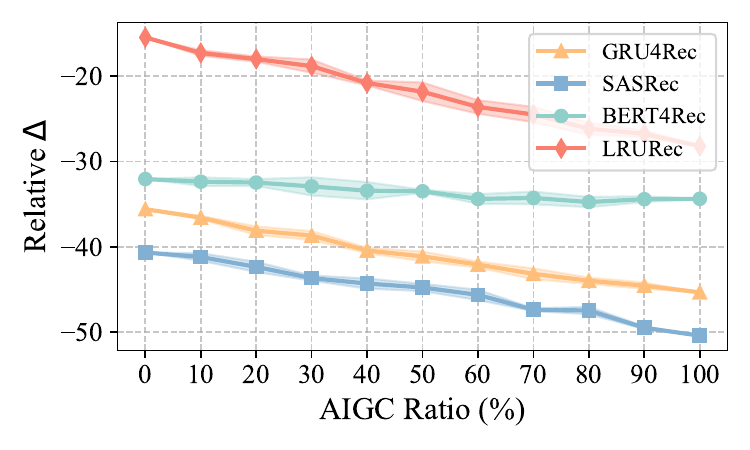}
    \label{fig:loop2_Amazon_Sports}
   }
   \vspace{-15pt}
  \caption{Relative $\Delta$ of recommendation models is depicted along with its $95\%$ confidence interval, shown with error bars. X-axis represents the ratio of AIGC in users' interaction sequence.}
  \label{fig:loop2}
\end{figure*}

\subsection{Source Bias in HGC-AIGC Coexist Phase}\label{sec:sec_loop2}

In this subsection, we validate the recommendation models during the HGC-AIGC coexist phase, which aims to explore whether preference will be amplified with the number of users' interaction on AIGC. When AIGC is further integrated into the recommender systems, users will interact with both HGC and AIGC. These items will be added to users' interaction history sequences, influencing the output of the recommendation models. In order to simulate this process, we train recommendation models on each dataset using items from $\mathcal{I}^H$. When testing, we vary the proportion of AIGC in users' interaction history sequence $s=\{i_1,\cdots,i_n \}$. For $i_t \in s$, it originates from $\mathcal{I}^H$ with probability $p$ and from $\mathcal{I}^G$ with probability $1-p$ where $p$ ranges from $0$ to $1$ in intervals of $0.1$. This allows us to simulate the impact of users' interactions on AIGC on the preference at different levels of AIGC propagation.

The results, as shown in Figure~\ref{fig:loop2}, indicate that the preference for AIGC of all sequential recommendation models increases as the proportion of AIGC in the historical sequence increases across the three datasets.
While the extent of preference exhibited by the same model varies across different datasets, they all show the same trend: the more AIGC the user interact with, the more pronounced the preference phenomenon becomes in recommender systems. 

\begin{center}
\fcolorbox{gray!60}{gray!20}{
  \parbox{0.95\linewidth}{
    \textbf{Finding 2:} In the feedback loop, the more users interact with AIGC, the model will recommend more AIGC in Top-$K$ serving, thereby amplifying the preference.
  }
}
\end{center}

\subsection{Preference in AIGC Dominate Phase}\label{sec:sec_loop3}
In this subsection, we validate the recommendation models during the AIGC dominate phase, which aims to explore whether preference will be further amplified with AIGC items participating in model training with the feedback loop. When AIGC dominates the recommender ecosystem in the future, it will influence any stage of the feedback loop, namely Top-$K$ serving, interaction, and training as shown in Figure~\ref{fig:recommendation_framework}, corresponding to the candidate list $\mathcal{I}$, users' interaction history sequence $s$, and the model's training data $\mathcal{S}$. To investigate the changing trend of preference during the AIGC dominate phase, we will construct a realistic scenario involving users' interactions. In this scenario, users are more inclined to interact with items positioned higher for model training and testing rather than merely mixing the data proportions.

Specifically, we will first train recommendation models with items from $\mathcal{I}^H$ and use the trained models to simulate users' interactions on item sets $\mathcal{I}^H \cup \mathcal{I}^G$. To simulate users' behavior, the position-based click model (PBM)~\cite{ai2021unbiased, richardson2007predicting} is used, where a interaction is registered only when the item is viewed and is relevant. Here, $E=1$ indicates that an item is examined by a user. For each impression, the likelihood of examination is determined by the position in the list of candidate items ~\cite{zhao2023unbiased}:
\begin{equation}
P\big(E=1|\text{rank}(i)=k\big) = k^{-\eta},
\label{eq:click_score}
\end{equation} where $\eta$ represents the hyper-parameter that controls the severity of position bias, and $\text{rank}(i)$ is the rank position of item $i$ in the candidate item list. After obtaining users' interaction results, we use the proportion of users' interactions with AIGC to adjust the proportion of AIGC items in the interaction sequence $s$, as mentioned in Section~\ref{sec:sec_loop2}, and then retrain the model using the aforementioned simulated interaction results and the mixed historical sequence. In the above training process, we iterate 10 times, assessing the preference level in each model at each iteration. During testing, the proportion of AIGC in users' interaction history matches the proportion used in training for that iteration. The complete training process is provided in Algorithm~\ref{algo_feedback}.

\begin{figure*}[h]
  \subfigure[Health with LR]
    {
    \includegraphics[width=0.65\columnwidth]{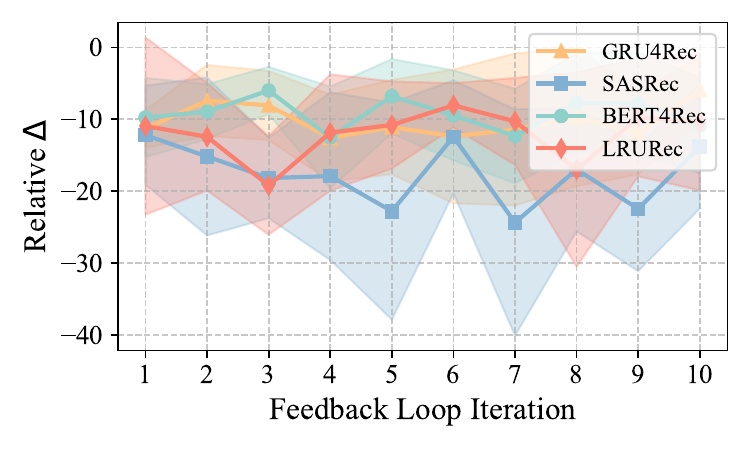}
    \label{fig:loop3_Amazon_Health_LR}
   }
  \subfigure[Health with NCM]
    {
    \includegraphics[width=0.65\columnwidth]{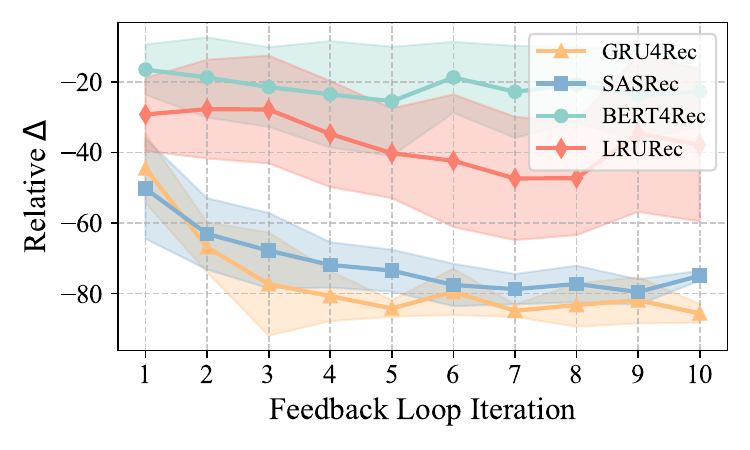}
    \label{fig:loop3_Amazon_Health_NCM}
   }
  \subfigure[Health with TNCM]
    {
    \includegraphics[width=0.65\columnwidth]{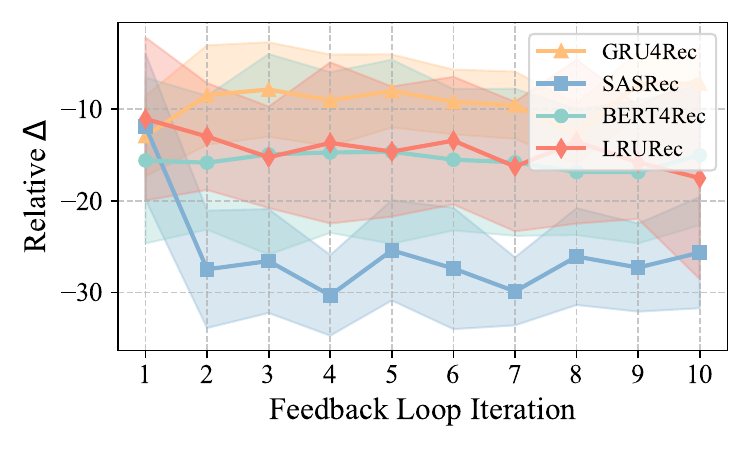}
    \label{fig:loop3_Amazon_Health_TNCM}
   }
   \vspace{-10pt}
  \caption{Comparison of Relative $\Delta$ for recommendation models under different click model settings across feedback loop iterations (X-axis), with $95\%$ confidence intervals represented by error bars.}
  \label{fig:loop3_more_click_model}
  \vspace{-5pt}
\end{figure*}

\begin{algorithm}[t]
\caption{Feedback Loop for Model Training}
\label{algo_feedback}
    \KwIn{Interaction dataset $\mathcal{S}$; number of feedback loop iterations $E$; parameters $\eta$, $p$}
    \KwOut{Trained models $f_{\theta}^1, f^2_{\theta}, \cdots f^E_{\theta}$}
    $p \leftarrow 0$ \\
    $\mathcal{S}^{e} \leftarrow \mathcal{S}$ \\
    \For{$e=1,\cdots,E$}
    {
        Train model $f^e_{\theta}$ on dataset $\mathcal{S}^{e}$ \\
        $\mathcal{S}^{e} \leftarrow \{ \} $ \\ 
        \For{$(s_t, i_{t+1})$ in $\mathcal{S}$}
        {
            $s_t \leftarrow \{i^{L} \mathbf{1}(\text{Bernoulli}(p^{e-1}) = 1) + i^{H} \mathbf{1}(\text{Bernoulli}(p^{e-1}) \neq 1) : i \in s_t\} $ \label{step1} \\

            Get users' interaction probabilities ${\mathcal{Y}}$ with Eq. \eqref{eq:click_score} \label{step2} \\
            Sample users' interaction item $i_{t+1}$ from $\mathcal{I}$ with $\mathcal{Y}$ \label{step3} \\
            
            $\mathcal{S}^e \leftarrow \mathcal{S}^e \cup (s_t, i_{t+1})$ \label{step4}\\
        }
        Update $p$ with probability of $i_{t+1}$ from $\mathcal{I}^G$ \label{step5} \\
    }

    \KwRet{$f_{\theta}^1, f_{\theta}^2, \cdots, f_{\theta}^E$} \\
\end{algorithm}

\begin{figure*}[t]
  \subfigure[Health]
    {
    \includegraphics[width=0.65\columnwidth]{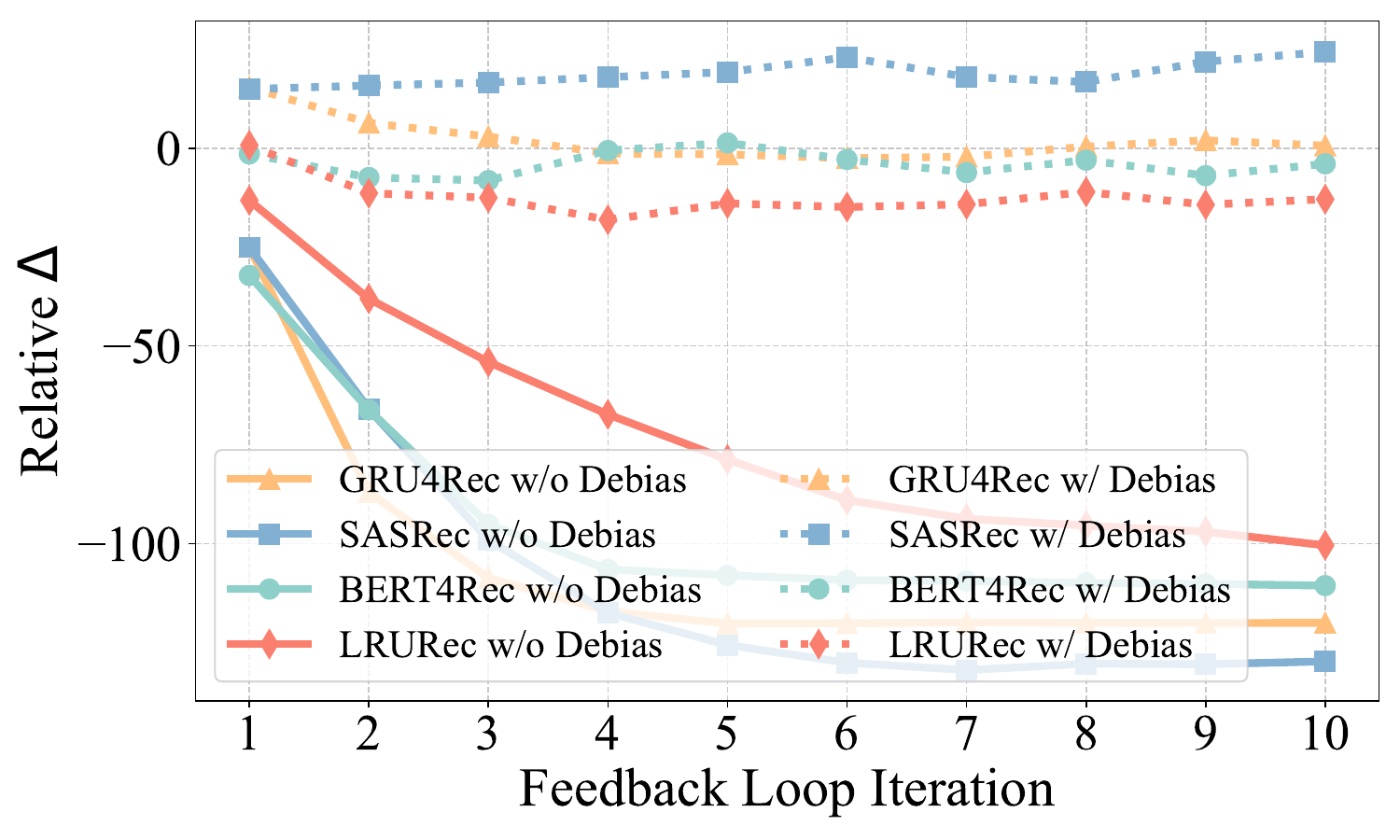}
    \label{fig:debias_Amazon_Health}
   }
  \subfigure[Beauty]
    {
    \includegraphics[width=0.65\columnwidth]{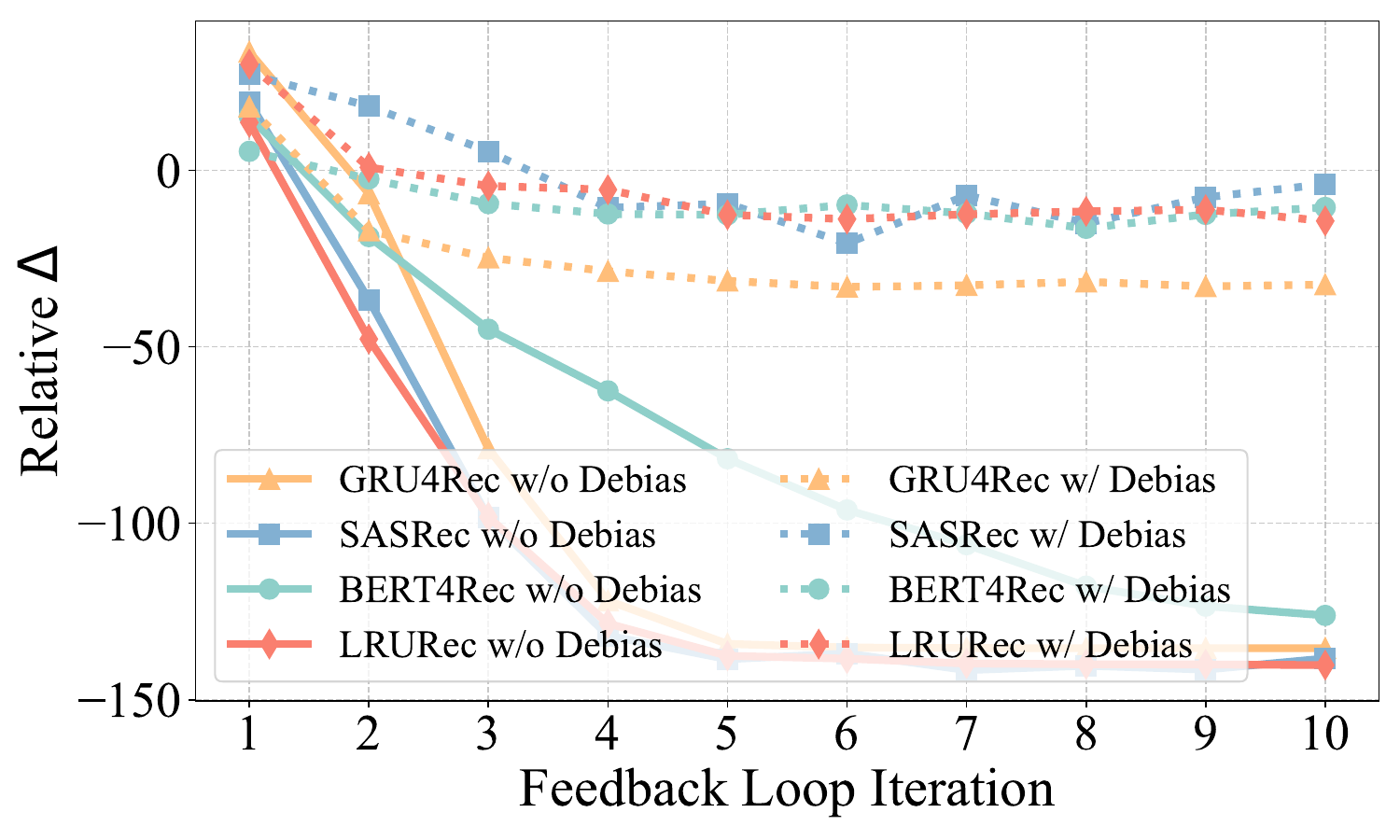}
    \label{fig:debias_Amazon_Beauty}
   }
  \subfigure[Sports]
    {
    \includegraphics[width=0.65\columnwidth]{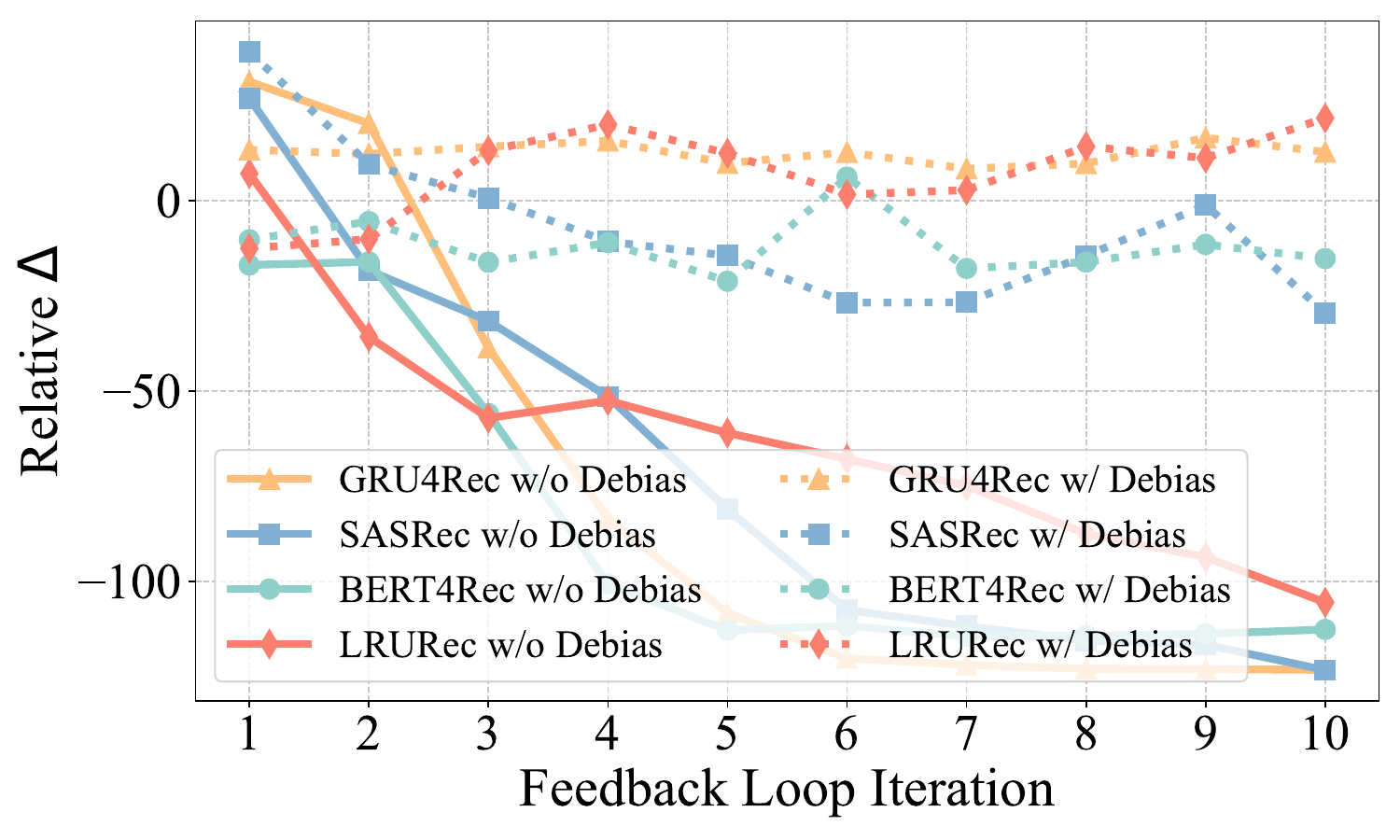}
    \label{fig:debias_Amazon_Sports}
   }
   \vspace{-15pt}
  \caption{Comparison of Relative $\Delta$ for recommendation models across feedback loop iterations (X-axis) on different datasets.}
  \vspace{-5pt}
  \label{fig:debias}
\end{figure*}

Under the condition of $\eta=+\infty$, we test the Relative $\Delta$ at different iterations of the feedback loop. 
The results in Figure~\ref{fig:debias} show that the absolute Relative $\Delta$ of all models increases with each iteration until it converges to a value near the end. This suggests that without intervening in the model's preference for AIGC, this preference will amplify with each feedback loop, ultimately leading to an AIGC-dominated content ecosystem. It is worth noting that on Beauty dataset, the models do not initially exhibit preference with Relative $\Delta > 0$. However, it still emerges and amplifies as the feedback loop progresses, further indicating the ubiquity of preference even if the model initially does not show a preference for AIGC. Additionally, we also record the performance changes of the model after the first loop and after the 20th loop. Results in Table~\ref{tab:feedback_performance} show that an excessively high proportion of AIGC not only disrupts the content ecosystem but also leads to a decline in model performance.

Although PBM is widely used and more realistic, we also implement three other training-based click models—LR, NCM, and TNCM~\cite{shirokikh2024neural}—for a more comprehensive evaluation. The results in Figure~\ref{fig:loop3_more_click_model} show similar phenomena as with PBM. Additionally, the model bias amplification through the feedback loop is more pronounced in these realistic neural models, emphasizing the inevitability of the issue. For simplicity and ease of comparison, we select PBM for the subsequent experiments.

\begin{center}
\fcolorbox{gray!60}{gray!20}{
  \parbox{0.95\linewidth}{
    \textbf{Finding 3:} Finally, by introducing AIGC pollute the feedback loop, including Top-$K$ serving, users' interactions, and model training, the preference will be pushed to the top.
  }
}
\end{center}

\section{Debias During the Feedback Loop}

In previous sections, we validate the presence of bias in recommender systems. Furthermore, with the proliferation of AIGC on the internet, bias amplifies throughout the feedback loop, thereby causing long-term impacts on the content ecosystem. Therefore, we need to eliminate the model's preference for AIGC. Although previous work~\cite{dai2024neural,xu2023ai} has attempted to address this bias, they do not account for the feedback loop inherent in real-world scenarios. In these settings, the margin loss used in their methods causes the model to ultimately favor HGC, leading to the collapse of the ecosystem with a dominance of HGC content. In this paper, we propose a new approach relying on L1 loss that effectively tackles this issue, ensuring a more stable and balanced content ecosystem.

Specifically, for each $i$ in $\mathcal{I}$, regardless of whether it originates from $\mathcal{I}^H$ or $\mathcal{I}^G$, its corresponding rewriting copy $i^{\prime}$ in $\mathcal{I}^{\prime}$ is derived from the rewriting process of LLM as described in Section~\ref{sec:data_construction}. In this way, we obtain the original training data triple $(s_t, i_{t+1}, i^{\prime}_{t+1})$ for feedback loop training. We utilize the L1 loss function to calculate the difference in scores between $i_{t+1}$ and $i^{\prime}_{t+1}$ as:
\begin{align}
\mathcal{L}_{\text{Debias-I}}=\sum_{s \in \mathcal{S}} \sum_{t=1}^{n-1} \left\lvert f_\theta(s_t, i_{t+1}^\prime) - f_\theta(s_t, i_{t+1}) \right\rvert,
\label{eq:item_debias}
\end{align}
which can eliminate the additional score introduced by the LLM rewriting process compared to the user interaction sequence $s$. Hence, it can be incorporated as a component of the loss function to alleviate the bias. What's more, for each item $i$ in the user interaction sequence $s$, we can obtain its rewritten copy $s^{\prime}$ by replacing each item $i$ with corresponding $i^{\prime}$. Again, we utilize the L1 loss function to calculate the difference in scores between $s$ and $s^{\prime}$ in comparison to candidate item $i$. Furthermore, in addition to aligning the embedding representations of user interaction sequences $s$ and $s^{\prime}$ before and after rewriting, we aim to minimize the entropy $\mathbb{H}$ of the embedding representation for each interaction sequence $s$ and $s^{\prime}$. This ensures that the embedding representations $\text{Emb}(s)$ generated from different $s$ composed of different items $i$ move farther away from each other. The debiasing loss for the history encoder side can be expressed as follows:
\begin{align}
    \mathcal{L}_{\text{Debias-U}} &=\sum_{s \in \mathcal{S}} \sum_{t=1}^{n-1} \left\lvert f_\theta(s^{\prime}_{t}, i_{t+1}) - f_\theta(s_{t}, i_{t+1}) \right\rvert \notag \\ &+ \mathbb{H}(\text{Softmax}(\text{Emb}(s^{\prime}_{t}))) + \mathbb{H}(\text{Softmax}(\text{Emb}(s_{t}))),
    \label{eq:user_debias}
\end{align}
which can measure the additional score resulting from the history encoder's preference for user interaction sequence $s^{\prime}$ combined with AIGC item, in comparison to item $i$. Therefore, this can also be used as part of the loss function to mitigate the bias caused by the history encoder. Based on the additional constraints defined in Eq.~\eqref{eq:item_debias} and Eq.~\eqref{eq:user_debias}, we can define the final loss for model training:
\begin{align}
    \mathcal{L} = \mathcal{L}_{\text{ranking}} + \alpha\mathcal{L}_{\text{Debias-I}} + \beta\mathcal{L}_{\text{Debias-U}},
    \label{eq:loss}
\end{align}
where $\mathcal{L}_{\text{ranking}}$ can be either contrastive loss or regression loss. $\alpha$ and $\beta$ are the debiasing coefficients that can balance the recommendation performance and the level of the bias. The larger coefficient indicates a greater penalty on the biased samples, which may result in a decrease in the recommendation performance.
%

\begin{table}[t]
\centering
\caption{Performance (NDCG@3) of recommendation models on different iteration of feedback loop. ``Iter=1'' and ``Iter=20'' indicate the 1st and 20th iterations of the feedback loop.}
\vspace{-5pt}
\label{tab:feedback_performance}
\resizebox{1.0\columnwidth}{!}{
\begin{tabular}{lccccccccccccc}
\hline\hline
 \multirow{2}{*}{Model} & \multicolumn{2}{c}{Health} & \multicolumn{2}{c}{Beauty} & \multicolumn{2}{c}{Sports} \\
\cmidrule(lr){2-3} \cmidrule(lr){4-5} \cmidrule(l){6-7}
 & Iter=1 & Iter=20 & Iter=1 & Iter=20 & Iter=1 & Iter=20 \\
\midrule
GRU4Rec  & 56.60 & 42.50 \worse{-14.10} & 60.18 & 43.73 \worse{-16.45} & 58.33 & 47.67 \worse{-10.66}  & \\
SASRec & 40.26 & 37.08 \worse{-3.18} & 50.53 & 36.44 \worse{-14.09} & 44.19 & 38.00 \worse{-6.19}  & \\
BERT4Rec  & 42.88 & 35.78 \worse{-7.10} & 42.92 & 35.76 \worse{-7.16} & 39.51 & 35.98 \worse{-3.53}  & \\
LRURec  & 43.00 & 37.85 \worse{-5.15} & 52.34 & 39.58 \worse{-12.76} & 50.64  & 44.28 \worse{-6.34} & \\
\hline\hline
\end{tabular}
}
\end{table}

\begin{table}[t]
\centering
\caption{Performance of recommendation models after feedback loop. ``\wo Debias'' refers to a model without our debiasing method, while ``\w Debias'' refers to one with it.}
\label{tab:bias_main}
\resizebox{1.0\columnwidth}{!}{
\begin{tabular}{lccccccccccccc}
\hline\hline
 \multirow{2}{*}{Model} & \multicolumn{2}{c}{Health} & \multicolumn{2}{c}{Beauty} & \multicolumn{2}{c}{Sports} \\
\cmidrule(lr){2-3} \cmidrule(lr){4-5} \cmidrule(l){6-7}
 & \wo Debias & Debias & \wo Debias & Debias & \wo Debias & Debias \\
\midrule
GRU4Rec  & 68.59 & 72.71 \better{+4.12} & 68.44 & 71.61 \better{+3.17} & 68.79 & 73.10 \better{+4.32} & \\
SASRec & 64.72 & 65.76 \better{+1.04} & 65.62 & 65.75 \better{+0.13} & 64.37 & 63.34 \worse{-1.03} & \\
BERT4Rec  & 63.82 & 63.28 \worse{-0.54} & 64.15 & 63.60 \worse{-0.55} & 62.20 & 60.55 \worse{-1.65} & \\
LRURec  & 65.37 & 66.73 \better{+1.36} & 66.59 & 68.31 \better{+1.72} & 64.60 & 64.63 \better{+0.03} & \\
\hline\hline
\end{tabular}
}
\end{table}

\subsection{Experimental Results}

Figure~\ref{fig:debias} illustrates the Relative $\Delta$ of the model and the debiasing model at different iterations of the feedback loop. The dashed line represents the model with our proposed method. Compared to previous methods, our approach focuses solely on the differences before and after rewriting, which enables us to continuously achieve debiasing effects in iterative feedback, resulting in a smaller absolute Relative $\Delta$. Furthermore, the dashed line remaining within a stable range also demonstrates that L1 loss can prevent the model from biasing towards either HGC or AIGC during the dynamic process.

Table~\ref{tab:bias_main} presents the ranking performance of models on all datasets . Our debiasing method not only eliminates biases but also enhances model performance in most cases. Suggested by these findings it can be found that introducing AIGC samples during the debiasing process of AIGC appears to enhance the model's capacity to differentiate between similar items.

\begin{figure}[t]  
    \centering    
    \includegraphics[width=1\linewidth]{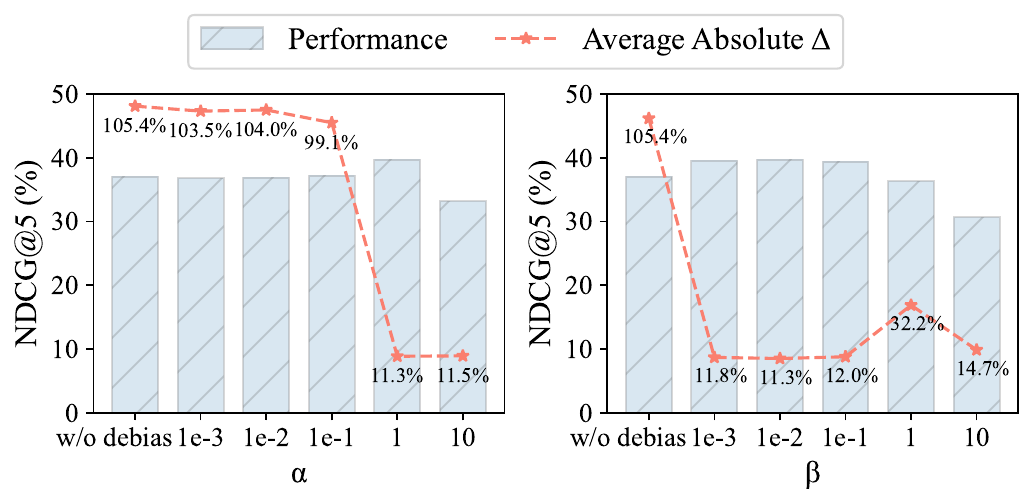}
    \vspace{-20pt}
    \caption{Performance and Average Absolute $\Delta$ of recommendation models with different coefficients $\alpha$ and $\beta$ in our proposed debiasing method.}
    \vspace{-10pt}
    \label{fig:tune_efficient}
\end{figure}

\subsection{Further Analysis}
\subsubsection{Performance w.r.t. the Coefficients $\alpha$ and $\beta$}
As shown in Eq.\eqref{eq:loss}, our debiasing method uses coefficients $\alpha$ and $\beta$ to balance the ranking loss and debiasing loss, achieving a trade-off between model performance and bias reduction. In the experiment, we vary $\alpha$ and $\beta$ within the range $\{1e\text{-}3, 1e\text{-}2, 1e\text{-}1, 1, 10\}$, while fixing the other coefficient at the value that yields the best recommendation performance. The Average Absolute $\Delta$ represents the average absolute value of the Relative $\Delta$ during the feedback loop. Models trained without debiasing constraints are labeled as ``\wo debias''.

The results on Figure~\ref{fig:tune_efficient} show that as $\alpha$ increases, the Average Absolute $\Delta$ decreases, indicating improved bias mitigation. The model also maintains ranking performance and outperforms the model without debiasing constraints. This improvement is likely due to the inclusion of AIGC samples, which may enhance the model's ability to distinguish relevance. However, when $\alpha$ becomes too large, performance declines, possibly because $\mathcal{L}_{\text{Debias-I}}$ shifts the focus too much on distinguishing HGC from AIGC, neglecting ranking. A similar trend is observed with $\beta$: increasing $\beta$ leads to performance degradation, likely because forcing interaction sequences to be closer disrupts the model's ranking capability.

\subsubsection{Ablation Study}
In this experiment, we investigate whether the two proposed components of loss, $\mathcal{L}_\text{Debias-U}$ and $\mathcal{L}_\text{Debias-I}$, can effectively eliminate the source bias. We conduct experiments to evaluate the Average Absolute $\Delta$ on the models trained only on our debiasing method without $\mathcal{L}_\text{Debias-U}$ and $\mathcal{L}_\text{Debias-I}$, denoted as ``\wo Debias-U'' and ``\wo Debias-I'', respectively. The results in Figure~\ref{fig:ablation} show that the Average Absolute $\Delta$ of the model improves across all models. After removing all debiasing constraints except for the ``\wo Debias-U model'' with SASRec implementation, the Average Absolute $\Delta$ increases. This observation confirms the effectiveness of constraining the item encoder and user encoder in our proposed loss function. Meanwhile, the $\mathcal{L}_{\text{Debais-I}}$ loss is more effective compared to the $\mathcal{L}_{\text{Debais-U}}$ loss may result from the fact that debiasing directly on the items used for evaluation is more straightforward.

\begin{table}[t]
\centering
\caption{Relative $\Delta$ on Health with varying AIGC settings: "ChatGPT" refers to AIGC generated solely by ChatGPT. "Mixed" refers to AIGC from multiple LLMs, with Llama3 used for debiased texts in both "Mixed" and "ChatGPT."}
\label{tab:realistic_setting}
\resizebox{1.0\columnwidth}{!}{
\begin{tabular}{lcccccc}
\hline\hline
 \multirow{3}{*}{Model} & \multicolumn{3}{c}{NDCG@5} & \multicolumn{3}{c}{MAP@5} \\
\cmidrule(lr){2-4} \cmidrule(lr){5-7} &
\wo Debias & Mixed & ChatGPT & \wo Debias & Mixed & ChatGPT \\
\midrule
GRU4Rec & -121.93 & -4.66 & 2.10   & -138.23 & -6.44 & 3.38   \\
SASRec & -122.11 & 5.19  & -29.32 & -136.92 & 7.74  & -35.57 \\
BERT4Rec & -109.10 & 12.84 & 15.17  & -123.43 & 15.08 & 18.23  \\
LRURec & -120.44 & -7.80 & -25.93 & -135.31 & -7.22 & -30.93 \\
\hline \hline
\end{tabular}
}
\end{table}

\begin{table}[t]
\centering
\caption{Average Absolute $\Delta$ of recommendation models with various debiasing method variants. ``\wo Debias-U'' refers to the model trained without $\mathcal{L}_{\text{Debias-U}}$, while ``\wo Deias-I'' refers to the model trained without $\mathcal{L}_{\text{Debias-I}}$.}
\label{fig:ablation}
\resizebox{1\columnwidth}{!}{
\begin{tabular}{ccccccc}
\hline\hline
 \multirow{3}{*}{Model} & \multicolumn{3}{c}{NDCG@5} & \multicolumn{3}{c}{MAP@5} \\
\cmidrule(lr){2-4} \cmidrule(lr){5-7}
 & Debias & \wo Debias-U & \wo Debias-I & Debias & \wo Debias-U & \wo Debias-I \\
\midrule
GRU4Rec & 6.45 & 8.07 \better{+1.62} & 130.00 \better{+123.55} & 7.12 & 9.65 \better{+2.53} & 142.89 \better{+135.77} \\
SASRec & 21.69 & 18.43 \worse{-3.26} & 124.80 \better{+103.11} & 24.39 & 20.84 \worse{-3.55} & 135.84 \better{+115.00} \\
BERT4Rec & 8.18 & 8.65 \better{+0.47} & 25.86 \better{+17.68} & 8.28 & 9.84 \better{+1.56} & 29.40 \better{+19.56} \\
LRURec & 8.97 & 29.72 \better{+20.75} & 122.30 \better{+113.33} & 11.34 & 32.74 \better{+21.4} & 131.93 \better{+120.59} \\
\hline \hline
\end{tabular}
}
\vspace{-10pt}
\end{table}

\subsubsection{More Realistic Debiasing Setting} In the previous setting, the AIGC used in the experiments is generated by ChatGPT. However, in real-world scenarios, there are many different types of LLMs. To better validate the effectiveness of our debiasing method in a real-world setting, we will use ChatGPT, Llama, Mistral, and Gemini to generate AIGC, and Llama3 to rewrite the text for debiasing (``Mixed'' setting ). We also use AIGC texts all generated by ChatGPT and use Llama3 for debiasing to simulate an extreme scenario (``ChatGPT'' setting). In Table~\ref{tab:realistic_setting}, whether AIGC is generated using a mix of LLMs or solely ChatGPT, after applying our debiasing method, the model achieves a smaller absolute value of Relative $\Delta$ compared to over 100 before debiasing. This means that a single LLM can correct biases in complex environments with AIGC generated by multiple LLMs. Moreover, our method only requires a one-time rewrite of the corpus by LLM and computation of AIGC-copy embeddings by PLM during training, with no additional cost during inference, making it practical for real-world use.

\subsubsection{Visualize of Interaction Sequence Representation}
\begin{figure}[t]
  \subfigure[\wo $\mathbb{H}(\text{Emb}(s))$ Maximized]
    {
    \includegraphics[width=0.47\columnwidth]{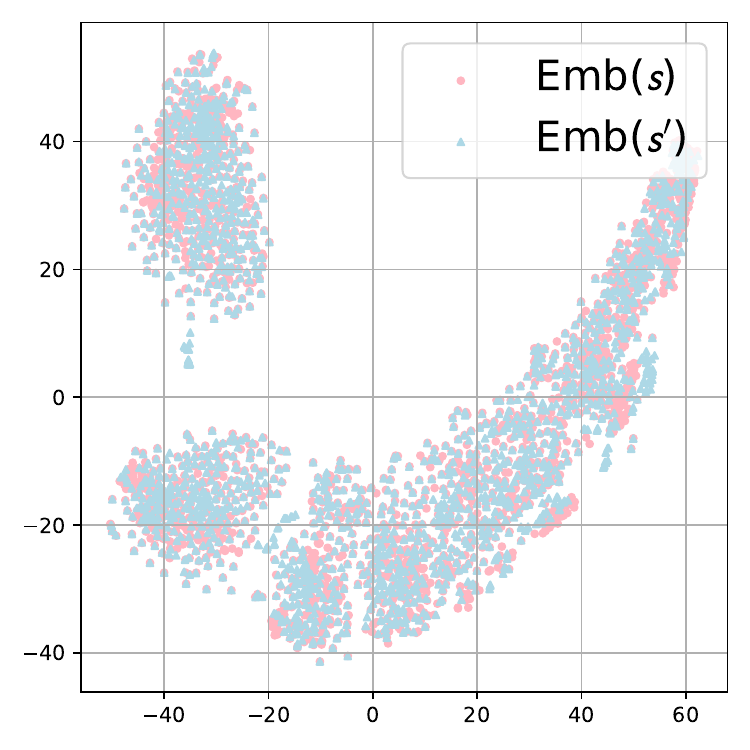}
    \label{fig:tsnet1}
   }
  \subfigure[\w $\mathbb{H}(\text{Emb}(s))$ Maximized]
    {
    \includegraphics[width=0.47\columnwidth]{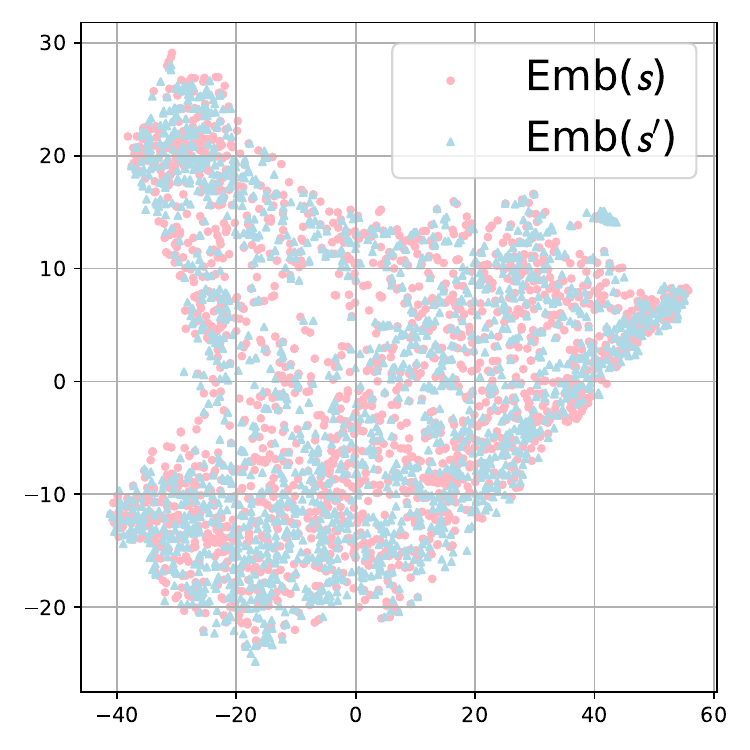}
    \label{fig:tsnet2}
   }
   \vspace{-10pt}
  \caption{User history embedding visualization of GRU4Rec trained without and with $\mathbb{H}(\text{Emb}(s))$ on Health dataset.}
  \vspace{-10pt}
  \label{fig:tsnet}
\end{figure}

As shown in Figure~\ref{fig:tsnet}, we visualize the interaction sequence representation $\textbf{Emb}(s)$ of models with various debiasing constraints using T-SNE~\cite{van2008visualizing}, in which the models are denoted as $\wo~\mathbb{H}(\text{Emb}(s))$ and $w/~\mathbb{H}(\text{Emb}(s))$ to indicate whether the term maximizing user embedding entropy is included in $\mathcal{L}_\text{Debias-U}$. Both types of debiasing constraints on $\text{Emb}(s)$ can maintain the mapping representation of historical sequences before and after rewriting.
However, our proposed debiasing constraints—minimizing both entropy $\mathbb{H}$ and the distance between $s$ and $s^\prime$—encourage a more uniform distribution of user history embeddings. This prevents different histories from collapsing into the same representation, preserving ranking performance while aligning AIGC and HGC sequences.

\section{Related Work}
\textbf{Large Language Models for Recommender Systems.}
Recent advancements in LLMs have attracted considerable interest among researchers to leverage these models~\cite{ wu2023survey,lin2023can,fan2023recommender,li2024survey} to develop an enhanced recommender system. Some works utilize LLMs to generate knowledge-rich texts or use LLM-derived embeddings to enhance recommender systems, known as LLM-enhanced recommender systems~\cite{xi2023towards, ren2023representation, wei2024llmrec}. Another line of work leverages LLMs that act as the ranking model to approach recommendation tasks, known as LLM-as-recommenders~\cite{hou2024large, dai2023uncovering, bao2023tallrec}. In addition to exploring how recommender systems can benefit from LLMs, we also need to consider the potential challenges that the development of LLMs may pose to recommender systems~\cite{dai2024bias, dai2025unifying}. Distinguished from these works, our study primarily investigates the impact of AIGC content on recommender systems, specifically focusing on the changes and influences of source bias in the feedback loop of recommender systems.

\noindent\textbf{Effects of Artificial Intelligence Generated Content}.
The rise of large language models (LLMs) has accelerated the spread of AIGC (AI-generated content), bringing broad societal and technological impacts~\cite{cao2023comprehensive,dai2024bias,wu2023ai,dai2025unifying}. AIGC raises concerns such as misinformation~\cite{chen2023combating}, harmful content~\cite{huang2023catastrophic}, and even performance degradation in future models when used for training~\cite{shumailov2023model,alemohammad2023self,briesch2023large}. Recent studies also reveal that neural retrieval models tend to favor AIGC, ranking it higher in text~\cite{dai2024neural,cocktail}, image~\cite{xu2023ai}, and video retrieval~\cite{gao2025generative}—a phenomenon known as source bias. \citet{wang2025perplexity} attributes this to the lower perplexity of AIGC, which aligns better with PLM-based retrieval models. While most work addresses this bias on the retrieval side, \citet{dai2025mitigating} instead leverages retriever feedback to construct preference data for LLM debiasing. In contrast, our study explores source bias in recommender systems, where AIGC affects not only model outputs and user behavior, but also future training data. This forms a feedback loop that can reinforce and amplify the bias over time.

\section{Conclusion}\label{sec:conclusion}
In this paper, we delve into exploring the effect of AIGC in recommender systems. Through extensive experiments with several representative recommendation models across three datasets from different domains, we uncover the prevalence of preference for AIGC in recommender systems. Furthermore, we validate that the preference is gradually amplified in the feedback loop, where AIGC will be incorporated into users' interaction histories and the training data as time progresses. To mitigate preference and prevent its further amplification in the feedback loop, we propose a black-box debiasing solution that ensures the impartiality of the model prediction towards both HGC and AIGC in the feedback loop. 

\begin{acks}
This work was funded by the National Key R\&D Program of China (2023YFA1008704), the National Natural Science Foundation of China (No. 62472426, 62276248),  Engineering Research Center of Next-Generation Intelligent Search and Recommendation, Ministry of Education, Major Innovation \& Planning Interdisciplinary Platform for the ``Double-First Class'' Initiative, fund for building world-class universities (disciplines) of Renmin University of China and the Youth Innovation Promotion Association CAS under Grants No.2023111. 
This work was supported by the Fundamental Research Funds for the Central Universities, and the Research Funds of Renmin University of China (RUC24QSDL013).
\end{acks}
\balance
\bibliographystyle{ACM-Reference-Format}
\bibliography{mybib}


\begin{thebibliography}{48}


\ifx \showCODEN    \undefined \def \showCODEN     #1{\unskip}     \fi
\ifx \showISBNx    \undefined \def \showISBNx     #1{\unskip}     \fi
\ifx \showISBNxiii \undefined \def \showISBNxiii  #1{\unskip}     \fi
\ifx \showISSN     \undefined \def \showISSN      #1{\unskip}     \fi
\ifx \showLCCN     \undefined \def \showLCCN      #1{\unskip}     \fi
\ifx \shownote     \undefined \def \shownote      #1{#1}          \fi
\ifx \showarticletitle \undefined \def \showarticletitle #1{#1}   \fi
\ifx \showURL      \undefined \def \showURL       {\relax}        \fi
\providecommand\bibfield[2]{#2}
\providecommand\bibinfo[2]{#2}
\providecommand\natexlab[1]{#1}
\providecommand\showeprint[2][]{arXiv:#2}

\bibitem[Ai et~al\mbox{.}(2021)]%
        {ai2021unbiased}
\bibfield{author}{\bibinfo{person}{Qingyao Ai}, \bibinfo{person}{Tao Yang}, \bibinfo{person}{Huazheng Wang}, {and} \bibinfo{person}{Jiaxin Mao}.} \bibinfo{year}{2021}\natexlab{}.
\newblock \showarticletitle{Unbiased learning to rank: online or offline?}
\newblock \bibinfo{journal}{\emph{ACM Transactions on Information Systems (TOIS)}} \bibinfo{volume}{39}, \bibinfo{number}{2} (\bibinfo{year}{2021}), \bibinfo{pages}{1--29}.
\newblock


\bibitem[Alemohammad et~al\mbox{.}(2023)]%
        {alemohammad2023self}
\bibfield{author}{\bibinfo{person}{Sina Alemohammad}, \bibinfo{person}{Josue Casco-Rodriguez}, \bibinfo{person}{Lorenzo Luzi}, \bibinfo{person}{Ahmed~Imtiaz Humayun}, \bibinfo{person}{Hossein Babaei}, \bibinfo{person}{Daniel LeJeune}, \bibinfo{person}{Ali Siahkoohi}, {and} \bibinfo{person}{Richard~G Baraniuk}.} \bibinfo{year}{2023}\natexlab{}.
\newblock \showarticletitle{Self-consuming generative models go mad}.
\newblock \bibinfo{journal}{\emph{arXiv preprint arXiv:2307.01850}}  \bibinfo{volume}{4} (\bibinfo{year}{2023}), \bibinfo{pages}{14}.
\newblock


\bibitem[Bahak et~al\mbox{.}(2023)]%
        {tan2023evaluation}
\bibfield{author}{\bibinfo{person}{Hossein Bahak}, \bibinfo{person}{Farzaneh Taheri}, \bibinfo{person}{Zahra Zojaji}, {and} \bibinfo{person}{Arefeh Kazemi}.} \bibinfo{year}{2023}\natexlab{}.
\newblock \showarticletitle{Evaluating chatgpt as a question answering system: A comprehensive analysis and comparison with existing models}.
\newblock \bibinfo{journal}{\emph{arXiv preprint arXiv:2312.07592}} (\bibinfo{year}{2023}).
\newblock


\bibitem[Bao et~al\mbox{.}(2023)]%
        {bao2023tallrec}
\bibfield{author}{\bibinfo{person}{Keqin Bao}, \bibinfo{person}{Jizhi Zhang}, \bibinfo{person}{Yang Zhang}, \bibinfo{person}{Wenjie Wang}, \bibinfo{person}{Fuli Feng}, {and} \bibinfo{person}{Xiangnan He}.} \bibinfo{year}{2023}\natexlab{}.
\newblock \showarticletitle{Tallrec: An effective and efficient tuning framework to align large language model with recommendation}. In \bibinfo{booktitle}{\emph{Proceedings of the 17th ACM Conference on Recommender Systems}}. \bibinfo{pages}{1007--1014}.
\newblock


\bibitem[Briesch et~al\mbox{.}(2023)]%
        {briesch2023large}
\bibfield{author}{\bibinfo{person}{Martin Briesch}, \bibinfo{person}{Dominik Sobania}, {and} \bibinfo{person}{Franz Rothlauf}.} \bibinfo{year}{2023}\natexlab{}.
\newblock \showarticletitle{Large language models suffer from their own output: An analysis of the self-consuming training loop}.
\newblock \bibinfo{journal}{\emph{arXiv preprint arXiv:2311.16822}} (\bibinfo{year}{2023}).
\newblock


\bibitem[Cao et~al\mbox{.}(2023)]%
        {cao2023comprehensive}
\bibfield{author}{\bibinfo{person}{Yihan Cao}, \bibinfo{person}{Siyu Li}, \bibinfo{person}{Yixin Liu}, \bibinfo{person}{Zhiling Yan}, \bibinfo{person}{Yutong Dai}, \bibinfo{person}{Philip~S Yu}, {and} \bibinfo{person}{Lichao Sun}.} \bibinfo{year}{2023}\natexlab{}.
\newblock \showarticletitle{A comprehensive survey of ai-generated content (aigc): A history of generative ai from gan to chatgpt}.
\newblock \bibinfo{journal}{\emph{arXiv preprint arXiv:2303.04226}} (\bibinfo{year}{2023}).
\newblock


\bibitem[Chen and Shu(2024)]%
        {chen2023combating}
\bibfield{author}{\bibinfo{person}{Canyu Chen} {and} \bibinfo{person}{Kai Shu}.} \bibinfo{year}{2024}\natexlab{}.
\newblock \showarticletitle{Combating misinformation in the age of llms: Opportunities and challenges}.
\newblock \bibinfo{journal}{\emph{AI Magazine}} \bibinfo{volume}{45}, \bibinfo{number}{3} (\bibinfo{year}{2024}), \bibinfo{pages}{354--368}.
\newblock


\bibitem[Dai et~al\mbox{.}(2024a)]%
        {cocktail}
\bibfield{author}{\bibinfo{person}{Sunhao Dai}, \bibinfo{person}{Weihao Liu}, \bibinfo{person}{Yuqi Zhou}, \bibinfo{person}{Liang Pang}, \bibinfo{person}{Rongju Ruan}, \bibinfo{person}{Gang Wang}, \bibinfo{person}{Zhenhua Dong}, \bibinfo{person}{Jun Xu}, {and} \bibinfo{person}{Ji-Rong Wen}.} \bibinfo{year}{2024}\natexlab{a}.
\newblock \showarticletitle{Cocktail: A Comprehensive Information Retrieval Benchmark with LLM-Generated Documents Integration}.
\newblock \bibinfo{journal}{\emph{Findings of the Association for Computational Linguistics: ACL 2024}} (\bibinfo{year}{2024}).
\newblock


\bibitem[Dai et~al\mbox{.}(2023)]%
        {dai2023uncovering}
\bibfield{author}{\bibinfo{person}{Sunhao Dai}, \bibinfo{person}{Ninglu Shao}, \bibinfo{person}{Haiyuan Zhao}, \bibinfo{person}{Weijie Yu}, \bibinfo{person}{Zihua Si}, \bibinfo{person}{Chen Xu}, \bibinfo{person}{Zhongxiang Sun}, \bibinfo{person}{Xiao Zhang}, {and} \bibinfo{person}{Jun Xu}.} \bibinfo{year}{2023}\natexlab{}.
\newblock \showarticletitle{Uncovering chatgpt’s capabilities in recommender systems}. In \bibinfo{booktitle}{\emph{Proceedings of the 17th ACM Conference on Recommender Systems}}. \bibinfo{pages}{1126--1132}.
\newblock


\bibitem[Dai et~al\mbox{.}(2024b)]%
        {dai2024bias}
\bibfield{author}{\bibinfo{person}{Sunhao Dai}, \bibinfo{person}{Chen Xu}, \bibinfo{person}{Shicheng Xu}, \bibinfo{person}{Liang Pang}, \bibinfo{person}{Zhenhua Dong}, {and} \bibinfo{person}{Jun Xu}.} \bibinfo{year}{2024}\natexlab{b}.
\newblock \showarticletitle{Bias and Unfairness in Information Retrieval Systems: New Challenges in the LLM Era}.
\newblock  (\bibinfo{year}{2024}), \bibinfo{pages}{6437--6447}.
\newblock


\bibitem[Dai et~al\mbox{.}(2025a)]%
        {dai2025unifying}
\bibfield{author}{\bibinfo{person}{Sunhao Dai}, \bibinfo{person}{Chen Xu}, \bibinfo{person}{Shicheng Xu}, \bibinfo{person}{Liang Pang}, \bibinfo{person}{Zhenhua Dong}, {and} \bibinfo{person}{Jun Xu}.} \bibinfo{year}{2025}\natexlab{a}.
\newblock \showarticletitle{Unifying Bias and Unfairness in Information Retrieval: New Challenges in the LLM Era}. In \bibinfo{booktitle}{\emph{Proceedings of the Eighteenth ACM International Conference on Web Search and Data Mining}}. \bibinfo{pages}{998–1001}.
\newblock


\bibitem[Dai et~al\mbox{.}(2025b)]%
        {dai2025mitigating}
\bibfield{author}{\bibinfo{person}{Sunhao Dai}, \bibinfo{person}{Yuqi Zhou}, \bibinfo{person}{Liang Pang}, \bibinfo{person}{Zhuoyang Li}, \bibinfo{person}{Zhaocheng Du}, \bibinfo{person}{Gang Wang}, {and} \bibinfo{person}{Jun Xu}.} \bibinfo{year}{2025}\natexlab{b}.
\newblock \showarticletitle{Mitigating Source Bias with LLM Alignment}. In \bibinfo{booktitle}{\emph{Proceedings of the 48th International ACM SIGIR Conference on Research and Development in Information Retrieval}}.
\newblock


\bibitem[Dai et~al\mbox{.}(2024c)]%
        {dai2024neural}
\bibfield{author}{\bibinfo{person}{Sunhao Dai}, \bibinfo{person}{Yuqi Zhou}, \bibinfo{person}{Liang Pang}, \bibinfo{person}{Weihao Liu}, \bibinfo{person}{Xiaolin Hu}, \bibinfo{person}{Yong Liu}, \bibinfo{person}{Xiao Zhang}, \bibinfo{person}{Gang Wang}, {and} \bibinfo{person}{Jun Xu}.} \bibinfo{year}{2024}\natexlab{c}.
\newblock \showarticletitle{Neural Retrievers are Biased Towards LLM-Generated Content}.
\newblock \bibinfo{journal}{\emph{Proceedings of the 30th ACM SIGKDD Conference on Knowledge Discovery and Data Mining}} (\bibinfo{year}{2024}).
\newblock


\bibitem[Devlin et~al\mbox{.}(2019)]%
        {kenton2019bert}
\bibfield{author}{\bibinfo{person}{Jacob Devlin}, \bibinfo{person}{Ming-Wei Chang}, \bibinfo{person}{Kenton Lee}, {and} \bibinfo{person}{Kristina Toutanova}.} \bibinfo{year}{2019}\natexlab{}.
\newblock \showarticletitle{Bert: Pre-training of deep bidirectional transformers for language understanding}. In \bibinfo{booktitle}{\emph{Proceedings of the 2019 conference of the North American chapter of the association for computational linguistics: human language technologies, volume 1 (long and short papers)}}. \bibinfo{pages}{4171--4186}.
\newblock


\bibitem[Gao et~al\mbox{.}(2025)]%
        {gao2025generative}
\bibfield{author}{\bibinfo{person}{Haowen Gao}, \bibinfo{person}{Liang Pang}, \bibinfo{person}{Shicheng Xu}, \bibinfo{person}{Leigang Qu}, \bibinfo{person}{Tat-Seng Chua}, \bibinfo{person}{Huawei Shen}, {and} \bibinfo{person}{Xueqi Cheng}.} \bibinfo{year}{2025}\natexlab{}.
\newblock \showarticletitle{Generative Ghost: Investigating Ranking Bias Hidden in AI-Generated Videos}.
\newblock \bibinfo{journal}{\emph{arXiv preprint arXiv:2502.07327}} (\bibinfo{year}{2025}).
\newblock


\bibitem[Hidasi et~al\mbox{.}(2015)]%
        {hidasi2015session}
\bibfield{author}{\bibinfo{person}{Bal{\'a}zs Hidasi}, \bibinfo{person}{Alexandros Karatzoglou}, \bibinfo{person}{Linas Baltrunas}, {and} \bibinfo{person}{Domonkos Tikk}.} \bibinfo{year}{2015}\natexlab{}.
\newblock \showarticletitle{Session-based recommendations with recurrent neural networks}.
\newblock \bibinfo{journal}{\emph{arXiv preprint arXiv:1511.06939}} (\bibinfo{year}{2015}).
\newblock


\bibitem[Hou et~al\mbox{.}(2024)]%
        {hou2024large}
\bibfield{author}{\bibinfo{person}{Yupeng Hou}, \bibinfo{person}{Junjie Zhang}, \bibinfo{person}{Zihan Lin}, \bibinfo{person}{Hongyu Lu}, \bibinfo{person}{Ruobing Xie}, \bibinfo{person}{Julian McAuley}, {and} \bibinfo{person}{Wayne~Xin Zhao}.} \bibinfo{year}{2024}\natexlab{}.
\newblock \showarticletitle{Large language models are zero-shot rankers for recommender systems}. In \bibinfo{booktitle}{\emph{European Conference on Information Retrieval}}. Springer, \bibinfo{pages}{364--381}.
\newblock


\bibitem[Huang et~al\mbox{.}(2025)]%
        {huang2023survey}
\bibfield{author}{\bibinfo{person}{Lei Huang}, \bibinfo{person}{Weijiang Yu}, \bibinfo{person}{Weitao Ma}, \bibinfo{person}{Weihong Zhong}, \bibinfo{person}{Zhangyin Feng}, \bibinfo{person}{Haotian Wang}, \bibinfo{person}{Qianglong Chen}, \bibinfo{person}{Weihua Peng}, \bibinfo{person}{Xiaocheng Feng}, \bibinfo{person}{Bing Qin}, {et~al\mbox{.}}} \bibinfo{year}{2025}\natexlab{}.
\newblock \showarticletitle{A survey on hallucination in large language models: Principles, taxonomy, challenges, and open questions}.
\newblock \bibinfo{journal}{\emph{ACM Transactions on Information Systems}} \bibinfo{volume}{43}, \bibinfo{number}{2} (\bibinfo{year}{2025}), \bibinfo{pages}{1--55}.
\newblock


\bibitem[Huang et~al\mbox{.}(2024)]%
        {huang2023catastrophic}
\bibfield{author}{\bibinfo{person}{Yangsibo Huang}, \bibinfo{person}{Samyak Gupta}, \bibinfo{person}{Mengzhou Xia}, \bibinfo{person}{Kai Li}, {and} \bibinfo{person}{Danqi Chen}.} \bibinfo{year}{2024}\natexlab{}.
\newblock \showarticletitle{Catastrophic Jailbreak of Open-source LLMs via Exploiting Generation}. In \bibinfo{booktitle}{\emph{12th International Conference on Learning Representations, ICLR 2024}}.
\newblock


\bibitem[Jiang et~al\mbox{.}(2023)]%
        {jiang2023mistral}
\bibfield{author}{\bibinfo{person}{Albert~Q Jiang}, \bibinfo{person}{Alexandre Sablayrolles}, \bibinfo{person}{Arthur Mensch}, \bibinfo{person}{Chris Bamford}, \bibinfo{person}{Devendra~Singh Chaplot}, \bibinfo{person}{Diego de~las Casas}, \bibinfo{person}{Florian Bressand}, \bibinfo{person}{Gianna Lengyel}, \bibinfo{person}{Guillaume Lample}, \bibinfo{person}{Lucile Saulnier}, {et~al\mbox{.}}} \bibinfo{year}{2023}\natexlab{}.
\newblock \showarticletitle{Mistral 7B}.
\newblock \bibinfo{journal}{\emph{arXiv}} (\bibinfo{year}{2023}).
\newblock


\bibitem[Kang and McAuley(2018)]%
        {kang2018self}
\bibfield{author}{\bibinfo{person}{Wang-Cheng Kang} {and} \bibinfo{person}{Julian McAuley}.} \bibinfo{year}{2018}\natexlab{}.
\newblock \showarticletitle{Self-attentive sequential recommendation}. In \bibinfo{booktitle}{\emph{2018 IEEE international conference on data mining (ICDM)}}. IEEE, \bibinfo{pages}{197--206}.
\newblock


\bibitem[Lai et~al\mbox{.}(2023)]%
        {lai2023chatgpt}
\bibfield{author}{\bibinfo{person}{Viet Lai}, \bibinfo{person}{Nghia Ngo}, \bibinfo{person}{Amir Pouran~Ben Veyseh}, \bibinfo{person}{Hiếu Mẫn}, \bibinfo{person}{Franck Dernoncourt}, \bibinfo{person}{Trung Bui}, {and} \bibinfo{person}{Thien Nguyen}.} \bibinfo{year}{2023}\natexlab{}.
\newblock \showarticletitle{ChatGPT Beyond English: Towards a Comprehensive Evaluation of Large Language Models in Multilingual Learning}. In \bibinfo{booktitle}{\emph{Findings of the Association for Computational Linguistics: EMNLP 2023}}. \bibinfo{pages}{13171--13189}.
\newblock


\bibitem[Li et~al\mbox{.}(2024)]%
        {li2024survey}
\bibfield{author}{\bibinfo{person}{Yongqi Li}, \bibinfo{person}{Xinyu Lin}, \bibinfo{person}{Wenjie Wang}, \bibinfo{person}{Fuli Feng}, \bibinfo{person}{Liang Pang}, \bibinfo{person}{Wenjie Li}, \bibinfo{person}{Liqiang Nie}, \bibinfo{person}{Xiangnan He}, {and} \bibinfo{person}{Tat-Seng Chua}.} \bibinfo{year}{2024}\natexlab{}.
\newblock \showarticletitle{A survey of generative search and recommendation in the era of large language models}.
\newblock \bibinfo{journal}{\emph{arXiv preprint arXiv:2404.16924}} (\bibinfo{year}{2024}).
\newblock


\bibitem[Lin et~al\mbox{.}(2025)]%
        {lin2023can}
\bibfield{author}{\bibinfo{person}{Jianghao Lin}, \bibinfo{person}{Xinyi Dai}, \bibinfo{person}{Yunjia Xi}, \bibinfo{person}{Weiwen Liu}, \bibinfo{person}{Bo Chen}, \bibinfo{person}{Hao Zhang}, \bibinfo{person}{Yong Liu}, \bibinfo{person}{Chuhan Wu}, \bibinfo{person}{Xiangyang Li}, \bibinfo{person}{Chenxu Zhu}, {et~al\mbox{.}}} \bibinfo{year}{2025}\natexlab{}.
\newblock \showarticletitle{How can recommender systems benefit from large language models: A survey}.
\newblock \bibinfo{journal}{\emph{ACM Transactions on Information Systems}} \bibinfo{volume}{43}, \bibinfo{number}{2} (\bibinfo{year}{2025}), \bibinfo{pages}{1--47}.
\newblock


\bibitem[Liu et~al\mbox{.}(2019)]%
        {liu2019roberta}
\bibfield{author}{\bibinfo{person}{Yinhan Liu}, \bibinfo{person}{Myle Ott}, \bibinfo{person}{Naman Goyal}, \bibinfo{person}{Jingfei Du}, \bibinfo{person}{Mandar Joshi}, \bibinfo{person}{Danqi Chen}, \bibinfo{person}{Omer Levy}, \bibinfo{person}{Mike Lewis}, \bibinfo{person}{Luke Zettlemoyer}, {and} \bibinfo{person}{Veselin Stoyanov}.} \bibinfo{year}{2019}\natexlab{}.
\newblock \showarticletitle{Roberta: A robustly optimized bert pretraining approach}.
\newblock \bibinfo{journal}{\emph{arXiv preprint arXiv:1907.11692}} (\bibinfo{year}{2019}).
\newblock


\bibitem[McAuley et~al\mbox{.}(2015)]%
        {mcauley2015image}
\bibfield{author}{\bibinfo{person}{Julian McAuley}, \bibinfo{person}{Christopher Targett}, \bibinfo{person}{Qinfeng Shi}, {and} \bibinfo{person}{Anton Van Den~Hengel}.} \bibinfo{year}{2015}\natexlab{}.
\newblock \showarticletitle{Image-based recommendations on styles and substitutes}. In \bibinfo{booktitle}{\emph{Proceedings of the 38th international ACM SIGIR conference on research and development in information retrieval}}. \bibinfo{pages}{43--52}.
\newblock


\bibitem[Ren et~al\mbox{.}(2024)]%
        {ren2023representation}
\bibfield{author}{\bibinfo{person}{Xubin Ren}, \bibinfo{person}{Wei Wei}, \bibinfo{person}{Lianghao Xia}, \bibinfo{person}{Lixin Su}, \bibinfo{person}{Suqi Cheng}, \bibinfo{person}{Junfeng Wang}, \bibinfo{person}{Dawei Yin}, {and} \bibinfo{person}{Chao Huang}.} \bibinfo{year}{2024}\natexlab{}.
\newblock \showarticletitle{Representation learning with large language models for recommendation}. In \bibinfo{booktitle}{\emph{Proceedings of the ACM Web Conference 2024}}. \bibinfo{pages}{3464--3475}.
\newblock


\bibitem[Richardson et~al\mbox{.}(2007)]%
        {richardson2007predicting}
\bibfield{author}{\bibinfo{person}{Matthew Richardson}, \bibinfo{person}{Ewa Dominowska}, {and} \bibinfo{person}{Robert Ragno}.} \bibinfo{year}{2007}\natexlab{}.
\newblock \showarticletitle{Predicting clicks: estimating the click-through rate for new ads}. In \bibinfo{booktitle}{\emph{Proceedings of the 16th international conference on World Wide Web}}. \bibinfo{pages}{521--530}.
\newblock


\bibitem[Schick(2020)]%
        {schick_deep_2020}
\bibfield{author}{\bibinfo{person}{Nina Schick}.} \bibinfo{year}{2020}\natexlab{}.
\newblock \bibinfo{booktitle}{\emph{Deep {Fakes} and the {Infocalypse}: {What} {You} {Urgently} {Need} {To} {Know}}}.
\newblock \bibinfo{publisher}{Monoray}.
\newblock


\bibitem[Shirokikh et~al\mbox{.}(2024)]%
        {shirokikh2024neural}
\bibfield{author}{\bibinfo{person}{Mikhail Shirokikh}, \bibinfo{person}{Ilya Shenbin}, \bibinfo{person}{Anton Alekseev}, \bibinfo{person}{Anna Volodkevich}, \bibinfo{person}{Alexey Vasilev}, \bibinfo{person}{Andrey~V Savchenko}, {and} \bibinfo{person}{Sergey Nikolenko}.} \bibinfo{year}{2024}\natexlab{}.
\newblock \showarticletitle{Neural Click Models for Recommender Systems}. In \bibinfo{booktitle}{\emph{Proceedings of the 47th International ACM SIGIR Conference on Research and Development in Information Retrieval}}. \bibinfo{pages}{2553--2558}.
\newblock


\bibitem[Shumailov et~al\mbox{.}(2023)]%
        {shumailov2023model}
\bibfield{author}{\bibinfo{person}{Ilia Shumailov}, \bibinfo{person}{Zakhar Shumaylov}, \bibinfo{person}{Yiren Zhao}, \bibinfo{person}{Yarin Gal}, \bibinfo{person}{Nicolas Papernot}, {and} \bibinfo{person}{Ross Anderson}.} \bibinfo{year}{2023}\natexlab{}.
\newblock \showarticletitle{Model dementia: Generated data makes models forget}.
\newblock \bibinfo{journal}{\emph{arXiv e-prints}} (\bibinfo{year}{2023}), \bibinfo{pages}{arXiv--2305}.
\newblock


\bibitem[Sun et~al\mbox{.}(2019)]%
        {sun2019bert4rec}
\bibfield{author}{\bibinfo{person}{Fei Sun}, \bibinfo{person}{Jun Liu}, \bibinfo{person}{Jian Wu}, \bibinfo{person}{Changhua Pei}, \bibinfo{person}{Xiao Lin}, \bibinfo{person}{Wenwu Ou}, {and} \bibinfo{person}{Peng Jiang}.} \bibinfo{year}{2019}\natexlab{}.
\newblock \showarticletitle{BERT4Rec: Sequential recommendation with bidirectional encoder representations from transformer}. In \bibinfo{booktitle}{\emph{Proceedings of the 28th ACM international conference on information and knowledge management}}. \bibinfo{pages}{1441--1450}.
\newblock


\bibitem[Team et~al\mbox{.}(2023)]%
        {team2023gemini}
\bibfield{author}{\bibinfo{person}{Gemini Team}, \bibinfo{person}{Rohan Anil}, \bibinfo{person}{Sebastian Borgeaud}, \bibinfo{person}{Jean-Baptiste Alayrac}, \bibinfo{person}{Jiahui Yu}, \bibinfo{person}{Radu Soricut}, \bibinfo{person}{Johan Schalkwyk}, \bibinfo{person}{Andrew~M Dai}, \bibinfo{person}{Anja Hauth}, \bibinfo{person}{Katie Millican}, {et~al\mbox{.}}} \bibinfo{year}{2023}\natexlab{}.
\newblock \showarticletitle{Gemini: a family of highly capable multimodal models}.
\newblock \bibinfo{journal}{\emph{arXiv preprint arXiv:2312.11805}} (\bibinfo{year}{2023}).
\newblock


\bibitem[Touvron et~al\mbox{.}(2023)]%
        {touvron2023llama}
\bibfield{author}{\bibinfo{person}{Hugo Touvron}, \bibinfo{person}{Louis Martin}, \bibinfo{person}{Kevin Stone}, \bibinfo{person}{Peter Albert}, \bibinfo{person}{Amjad Almahairi}, \bibinfo{person}{Yasmine Babaei}, \bibinfo{person}{Nikolay Bashlykov}, \bibinfo{person}{Soumya Batra}, \bibinfo{person}{Prajjwal Bhargava}, \bibinfo{person}{Shruti Bhosale}, {et~al\mbox{.}}} \bibinfo{year}{2023}\natexlab{}.
\newblock \showarticletitle{Llama 2: Open foundation and fine-tuned chat models}.
\newblock \bibinfo{journal}{\emph{arXiv preprint arXiv:2307.09288}} (\bibinfo{year}{2023}).
\newblock


\bibitem[Van~der Maaten and Hinton(2008)]%
        {van2008visualizing}
\bibfield{author}{\bibinfo{person}{Laurens Van~der Maaten} {and} \bibinfo{person}{Geoffrey Hinton}.} \bibinfo{year}{2008}\natexlab{}.
\newblock \showarticletitle{Visualizing data using t-SNE.}
\newblock \bibinfo{journal}{\emph{Journal of machine learning research}} \bibinfo{volume}{9}, \bibinfo{number}{11} (\bibinfo{year}{2008}).
\newblock


\bibitem[Wang et~al\mbox{.}(2025)]%
        {wang2025perplexity}
\bibfield{author}{\bibinfo{person}{Haoyu Wang}, \bibinfo{person}{Sunhao Dai}, \bibinfo{person}{Haiyuan Zhao}, \bibinfo{person}{Liang Pang}, \bibinfo{person}{Xiao Zhang}, \bibinfo{person}{Gang Wang}, \bibinfo{person}{Zhenhua Dong}, \bibinfo{person}{Jun Xu}, {and} \bibinfo{person}{Ji-Rong Wen}.} \bibinfo{year}{2025}\natexlab{}.
\newblock \showarticletitle{Perplexity Trap: PLM-Based Retrievers Overrate Low Perplexity Documents}. In \bibinfo{booktitle}{\emph{13th International Conference on Learning Representations, ICLR 2025}}.
\newblock


\bibitem[Wei et~al\mbox{.}(2024)]%
        {wei2024llmrec}
\bibfield{author}{\bibinfo{person}{Wei Wei}, \bibinfo{person}{Xubin Ren}, \bibinfo{person}{Jiabin Tang}, \bibinfo{person}{Qinyong Wang}, \bibinfo{person}{Lixin Su}, \bibinfo{person}{Suqi Cheng}, \bibinfo{person}{Junfeng Wang}, \bibinfo{person}{Dawei Yin}, {and} \bibinfo{person}{Chao Huang}.} \bibinfo{year}{2024}\natexlab{}.
\newblock \showarticletitle{Llmrec: Large language models with graph augmentation for recommendation}. In \bibinfo{booktitle}{\emph{Proceedings of the 17th ACM International Conference on Web Search and Data Mining}}. \bibinfo{pages}{806--815}.
\newblock


\bibitem[Wu et~al\mbox{.}(2023)]%
        {wu2023ai}
\bibfield{author}{\bibinfo{person}{Jiayang Wu}, \bibinfo{person}{Wensheng Gan}, \bibinfo{person}{Zefeng Chen}, \bibinfo{person}{Shicheng Wan}, {and} \bibinfo{person}{Hong Lin}.} \bibinfo{year}{2023}\natexlab{}.
\newblock \showarticletitle{Ai-generated content (aigc): A survey}.
\newblock \bibinfo{journal}{\emph{arXiv preprint arXiv:2304.06632}} (\bibinfo{year}{2023}).
\newblock


\bibitem[Wu et~al\mbox{.}(2024)]%
        {wu2023survey}
\bibfield{author}{\bibinfo{person}{Likang Wu}, \bibinfo{person}{Zhi Zheng}, \bibinfo{person}{Zhaopeng Qiu}, \bibinfo{person}{Hao Wang}, \bibinfo{person}{Hongchao Gu}, \bibinfo{person}{Tingjia Shen}, \bibinfo{person}{Chuan Qin}, \bibinfo{person}{Chen Zhu}, \bibinfo{person}{Hengshu Zhu}, \bibinfo{person}{Qi Liu}, {et~al\mbox{.}}} \bibinfo{year}{2024}\natexlab{}.
\newblock \showarticletitle{A survey on large language models for recommendation}.
\newblock \bibinfo{journal}{\emph{World Wide Web}} \bibinfo{volume}{27}, \bibinfo{number}{5} (\bibinfo{year}{2024}), \bibinfo{pages}{60}.
\newblock


\bibitem[Xi et~al\mbox{.}(2024)]%
        {xi2023towards}
\bibfield{author}{\bibinfo{person}{Yunjia Xi}, \bibinfo{person}{Weiwen Liu}, \bibinfo{person}{Jianghao Lin}, \bibinfo{person}{Xiaoling Cai}, \bibinfo{person}{Hong Zhu}, \bibinfo{person}{Jieming Zhu}, \bibinfo{person}{Bo Chen}, \bibinfo{person}{Ruiming Tang}, \bibinfo{person}{Weinan Zhang}, {and} \bibinfo{person}{Yong Yu}.} \bibinfo{year}{2024}\natexlab{}.
\newblock \showarticletitle{Towards open-world recommendation with knowledge augmentation from large language models}. In \bibinfo{booktitle}{\emph{Proceedings of the 18th ACM Conference on Recommender Systems}}. \bibinfo{pages}{12--22}.
\newblock


\bibitem[Xu et~al\mbox{.}(2024a)]%
        {xu2023ai}
\bibfield{author}{\bibinfo{person}{Shicheng Xu}, \bibinfo{person}{Danyang Hou}, \bibinfo{person}{Liang Pang}, \bibinfo{person}{Jingcheng Deng}, \bibinfo{person}{Jun Xu}, \bibinfo{person}{Huawei Shen}, {and} \bibinfo{person}{Xueqi Cheng}.} \bibinfo{year}{2024}\natexlab{a}.
\newblock \showarticletitle{Invisible relevance bias: Text-image retrieval models prefer ai-generated images}. In \bibinfo{booktitle}{\emph{Proceedings of the 47th international ACM SIGIR conference on research and development in information retrieval}}. \bibinfo{pages}{208--217}.
\newblock


\bibitem[Xu et~al\mbox{.}(2024b)]%
        {xu2024search}
\bibfield{author}{\bibinfo{person}{Shicheng Xu}, \bibinfo{person}{Liang Pang}, \bibinfo{person}{Huawei Shen}, \bibinfo{person}{Xueqi Cheng}, {and} \bibinfo{person}{Tat-Seng Chua}.} \bibinfo{year}{2024}\natexlab{b}.
\newblock \showarticletitle{Search-in-the-Chain: Interactively Enhancing Large Language Models with Search for Knowledge-intensive Tasks}. In \bibinfo{booktitle}{\emph{Proceedings of the ACM Web Conference 2024}}. \bibinfo{pages}{1362--1373}.
\newblock


\bibitem[Yue et~al\mbox{.}(2024)]%
        {yue2024linear}
\bibfield{author}{\bibinfo{person}{Zhenrui Yue}, \bibinfo{person}{Yueqi Wang}, \bibinfo{person}{Zhankui He}, \bibinfo{person}{Huimin Zeng}, \bibinfo{person}{Julian McAuley}, {and} \bibinfo{person}{Dong Wang}.} \bibinfo{year}{2024}\natexlab{}.
\newblock \showarticletitle{Linear recurrent units for sequential recommendation}. In \bibinfo{booktitle}{\emph{Proceedings of the 17th ACM international conference on web search and data mining}}. \bibinfo{pages}{930--938}.
\newblock


\bibitem[Zhang et~al\mbox{.}(2023)]%
        {zhang2023extractive}
\bibfield{author}{\bibinfo{person}{Haopeng Zhang}, \bibinfo{person}{Xiao Liu}, {and} \bibinfo{person}{Jiawei Zhang}.} \bibinfo{year}{2023}\natexlab{}.
\newblock \showarticletitle{Extractive Summarization via ChatGPT for Faithful Summary Generation}. In \bibinfo{booktitle}{\emph{Findings of the Association for Computational Linguistics: EMNLP 2023}}. \bibinfo{pages}{3270--3278}.
\newblock


\bibitem[Zhao et~al\mbox{.}(2023a)]%
        {zhao2023unbiased}
\bibfield{author}{\bibinfo{person}{Haiyuan Zhao}, \bibinfo{person}{Jun Xu}, \bibinfo{person}{Xiao Zhang}, \bibinfo{person}{Guohao Cai}, \bibinfo{person}{Zhenhua Dong}, {and} \bibinfo{person}{Ji-Rong Wen}.} \bibinfo{year}{2023}\natexlab{a}.
\newblock \showarticletitle{Unbiased Top-k Learning to Rank with Causal Likelihood Decomposition}. In \bibinfo{booktitle}{\emph{Proceedings of the Annual International ACM SIGIR Conference on Research and Development in Information Retrieval in the Asia Pacific Region}}. \bibinfo{pages}{129--138}.
\newblock


\bibitem[Zhao et~al\mbox{.}(2023b)]%
        {zhao2023survey}
\bibfield{author}{\bibinfo{person}{Wayne~Xin Zhao}, \bibinfo{person}{Kun Zhou}, \bibinfo{person}{Junyi Li}, \bibinfo{person}{Tianyi Tang}, \bibinfo{person}{Xiaolei Wang}, \bibinfo{person}{Yupeng Hou}, \bibinfo{person}{Yingqian Min}, \bibinfo{person}{Beichen Zhang}, \bibinfo{person}{Junjie Zhang}, \bibinfo{person}{Zican Dong}, {et~al\mbox{.}}} \bibinfo{year}{2023}\natexlab{b}.
\newblock \showarticletitle{A survey of large language models}.
\newblock \bibinfo{journal}{\emph{arXiv preprint arXiv:2303.18223}} \bibinfo{volume}{1}, \bibinfo{number}{2} (\bibinfo{year}{2023}).
\newblock


\bibitem[Zhao et~al\mbox{.}(2024)]%
        {fan2023recommender}
\bibfield{author}{\bibinfo{person}{Zihuai Zhao}, \bibinfo{person}{Wenqi Fan}, \bibinfo{person}{Jiatong Li}, \bibinfo{person}{Yunqing Liu}, \bibinfo{person}{Xiaowei Mei}, \bibinfo{person}{Yiqi Wang}, \bibinfo{person}{Zhen Wen}, \bibinfo{person}{Fei Wang}, \bibinfo{person}{Xiangyu Zhao}, \bibinfo{person}{Jiliang Tang}, {et~al\mbox{.}}} \bibinfo{year}{2024}\natexlab{}.
\newblock \showarticletitle{Recommender systems in the era of large language models (llms)}.
\newblock \bibinfo{journal}{\emph{IEEE Transactions on Knowledge and Data Engineering}} (\bibinfo{year}{2024}).
\newblock


\bibitem[Zhou and Zafarani(2020)]%
        {zhou2020survey}
\bibfield{author}{\bibinfo{person}{Xinyi Zhou} {and} \bibinfo{person}{Reza Zafarani}.} \bibinfo{year}{2020}\natexlab{}.
\newblock \showarticletitle{A survey of fake news: Fundamental theories, detection methods, and opportunities}.
\newblock \bibinfo{journal}{\emph{ACM Computing Surveys (CSUR)}} \bibinfo{volume}{53}, \bibinfo{number}{5} (\bibinfo{year}{2020}), \bibinfo{pages}{1--40}.
\newblock


\end{thebibliography}

\end{document}